\documentclass{ws-ijbc}

\usepackage{graphicx}
\usepackage{amsmath,amssymb}
\usepackage{bm}
\usepackage{booktabs}
\usepackage{multirow}
\usepackage{subcaption}
\begin{document}

\markboth{Z. Fu and U. Pillai}{Suppressing Chaos Without Silencing}

\title{Suppressing Chaos Without Silencing Excitable Oscillators}

\author{Zixuan Fu}
\address{Department of Electrical and Computer Engineering,\\
Tandon School of Engineering, New York University,\\
370 Jay Street, Brooklyn, NY 11201, USA\\
zf2375@nyu.edu}

\author{Unnikrishna Pillai}
\address{Department of Electrical and Computer Engineering,\\
Tandon School of Engineering, New York University,\\
370 Jay Street, Brooklyn, NY 11201, USA\\
up240@nyu.edu}

\maketitle


\begin{abstract}
A negative Lyapunov exponent is not a sufficient goal for controlling an
excitable oscillator. A gain can remove chaos by removing the excitation that
the model is meant to keep. This manuscript studies that failure mode in three
models: the forced FitzHugh--Nagumo neuron, the forced Aliev--Panfilov cardiac
cell, and the autonomous Hindmarsh--Rose neuron. Chaos is measured by the
largest state Lyapunov exponent. Excitation is measured by threshold crossings,
by the excursion of the fast variable, and by the firing rate. Recovery damping,
tracking of an unstable periodic orbit, Pyragas delayed feedback, and a change
of the fast time scale are compared on the same chaotic regimes. In the forced models, large
damping and large rate changes reach regularity by silencing the cell, while
tracking and delayed feedback keep firing. In the autonomous model, recovery
damping raises the firing rate several-fold instead. A classical matrix-measure
bound links these outcomes to the channel through which the gain acts. A constant-metric contraction certificate coexists with substantial firing only under tracking, which subtracts a gain from the fast direction.
\end{abstract}

\keywords{Chaos control; excitable systems; FitzHugh--Nagumo model;
Aliev--Panfilov model; Hindmarsh--Rose model; Lyapunov exponent; contraction
analysis; Pyragas control.}

\section{Introduction}\label{sec:intro}

This manuscript shows that suppressing chaos and keeping excitation are
separate goals in an excitable oscillator, and that the channel through which a
feedback gain acts decides whether both are met. An excitable model has an
activation threshold and a large excursion, the spike or action potential. A
gain can make the largest Lyapunov exponent negative by keeping the trajectory
away from that excursion. The controlled cell is then regular because it is
silent, depolarized, or reduced to a small oscillation. Loss of oscillation
under coupling is known as amplitude death or oscillation death
\cite{koseska2013}. This manuscript treats regularity and retained excitation
as two goals and measures both. Four interventions are compared on the same
chaotic regimes of the forced FitzHugh--Nagumo neuron, the forced
Aliev--Panfilov cardiac cell, and the autonomous Hindmarsh--Rose neuron, with
the same measures throughout.

An excitation index, an excursion ratio, and a firing-rate ratio are placed
next to the Lyapunov exponent. Maps over gain and drive built from these
measures separate regular responses that keep firing from regular responses
produced by silencing. In the two forced models, large recovery damping and
large rate scaling reach regularity by silencing the cell. Orbit tracking and
Pyragas delayed feedback reach regularity while the cell keeps firing. In the
autonomous model, recovery damping does not silence the cell at any scanned
gain. It raises the
firing rate several-fold instead.

A threshold calculation explains this comparison. In the FitzHugh--Nagumo
model, recovery damping leaves a positive frozen growth rate on the threshold
line at every gain. Fast-variable tracking removes that growth rate above an
explicit gain. For autonomous closed loops, a known contraction property is
stated in trajectory-wise form. A constant-metric certificate along a bounded
trajectory forces convergence to an equilibrium. A certificate therefore cannot
coexist with firing under recovery damping or rate scaling in the
Hindmarsh--Rose model. Numerically, a constant-metric contraction certificate
coexists with substantial firing only under tracking. Under rate scaling in the
Aliev--Panfilov model, it coexists with threshold crossings of small excursion.
None of the control laws is new, and the matrix-measure bound is classical.
The contribution lies in the excitation measures, the comparison on common
regimes, and the threshold calculation that explains the comparison.

Excitable oscillators are low-order models of cardiac and neural cells. Under
periodic forcing they show phase locking, period doubling, and chaos
\cite{guevara1981,glass,strogatz}. Periodically forced Bonhoeffer--van der Pol
oscillators, a form of the FitzHugh--Nagumo model, continue to show new
high-order mixed-mode oscillations \cite{sekikawa2026}. The Hindmarsh--Rose neuron bursts without any forcing \cite{hindmarsh1984}, and its chaotic regimes are well documented \cite{barrio2014}. A two-dimensional
Hindmarsh--Rose neuron coupled to a Hopfield neuron through a locally active
memristor also shows chaotic bursting \cite{huang2024}. Chaos control is a classical subject. OGY
control stabilizes an unstable periodic orbit with small parameter changes
\cite{ogy}. Pyragas control uses a delayed feedback that vanishes on the target
orbit \cite{pyragas}. Reviews and books cover the field
\cite{chendong1998,scholl2008}.

Linear state feedback with one gain is not a new control principle. The forced
Duffing oscillator was controlled with a single nonzero gain that tracks an
inherent unstable periodic orbit \cite{chendong1993}. That work used a weighted
energy functional and a high-gain argument. The tracking law used below is that
law, adapted to excitable models. It serves as a baseline.

Chaos control in excitable systems is also established. Cardiac chaos was
controlled in experiments \cite{garfinkel1992}, and cardiac alternans was
suppressed by dynamic control \cite{hall1997}. A model-independent technique
was developed for excitable physiological systems \cite{christini1997}.
Nonfeedback methods were compared in the FitzHugh--Nagumo equation
\cite{rajasekar1997}. Delayed feedback stabilized regular behavior in a paced
FitzHugh--Nagumo oscillator \cite{bleich2000}. That study also stabilized a
nonspiking orbit in the rapid-pacing regime. Spiking and nonspiking controlled orbits were therefore already distinguished in the delayed-feedback setting.
Filtered and phase feedback controlled a driven FitzHugh--Nagumo circuit in
spiking and nonspiking regimes \cite{ciszak2012,meucci2012}. The phase-feedback
study kept the spike amplitude within a few percent of the uncontrolled signal
\cite{meucci2012}. Chaos in the Hindmarsh--Rose model was controlled by state
feedback \cite{sabbagh2000} and by delayed self-feedback \cite{yu2009}. Spiral
chaos in the spatially extended Aliev--Panfilov model was eliminated by
periodic forcing \cite{sakaguchi2003}.
Feedback-timed pacing has also been studied in two-dimensional excitable-media
models of cardiac tissue \cite{suth2024}.

\emph{Chaos} means bounded aperiodic
dynamics with a positive largest state Lyapunov exponent. For the forced
models, the exponent is conditional on the prescribed forcing phase. For the
autonomous model, it is the ordinary state exponent. A complicated phase
portrait alone is not enough. \emph{Regularity} means that this exponent
is not positive within a stated numerical tolerance.
\emph{Firing} means repeated upward crossings of an activation threshold.
\emph{Silencing} means that such crossings stop. A crossing alone does not
prove a full-amplitude spike, so the excursion and the firing rate are reported
with it.

\section{Models, Interventions, and Measures}\label{sec:models}

Let $\bm x=(x_1,\ldots,x_m)^\top$ denote the state, where $m$ is the state
dimension and $x_1$ is the fast variable. The subscript $s$ denotes the slowest
recovery component $x_s$. Additive control inputs enter the fast and slowest
recovery equations as $c_1(t)$ and $c_s(t)$, respectively. The forced FitzHugh--Nagumo model \cite{fitzhugh} has state
$\bm x=(v,w)^\top$, with $x_1=v$ and $x_s=w$. The symbols $A$ and $\omega$
denote forcing amplitude and angular frequency. The symbols $\varepsilon$, $a$,
and $b$ are model parameters. The equations are
\begin{align}
\dot v &= v-\tfrac13 v^3-w+A\cos(\omega t)+c_1, \label{eq:fhn1}\\
\dot w &= \varepsilon\,(v+a-bw)+c_s. \label{eq:fhn2}
\end{align}
The reference parameter set is
\begin{equation}
(\varepsilon,a,b,\omega,A)=(0.1,0.7,0.8,1.15,1.30).
\label{eq:fhnpar}
\end{equation}
These values define the reference chaotic regime.

The forced Aliev--Panfilov model \cite{alievpanfilov} has state
$\bm x=(u,q)^\top$, with $x_1=u$ and $x_s=q$. The symbols $F$ and $\omega$
denote forcing amplitude and angular frequency. The symbols $K$, $\alpha$,
$\epsilon_0$, $\mu_1$, and $\mu_2$ are model parameters. The equations are
\begin{align}
\dot u &= -Ku(u-\alpha)(u-1)-uq+F\cos(\omega t)+c_1, \label{eq:ap1}\\
\dot q &= \epsilon(u,q)\,\big(-q-Ku(u-\alpha-1)\big)+c_s, \label{eq:ap2}\\
\epsilon(u,q) &= \epsilon_0+\frac{\mu_1 q}{u+\mu_2}. \label{eq:ap3}
\end{align}
The reference parameter set is
\begin{equation}
(K,\alpha,\epsilon_0,\mu_1,\mu_2,\omega,F)
=(8,0.12,0.01,0.2,0.3,1.8,0.40).
\label{eq:appar}
\end{equation}
Both forced models have forcing period
\begin{equation}
T=\frac{2\pi}{\omega}.
\label{eq:forcing-period}
\end{equation}

The Hindmarsh--Rose model \cite{hindmarsh1984} has state
$\bm x=(x,y,z)^\top$, with $x_1=x$ and $x_s=z$. The symbols $I$ and $\beta$
denote applied current and a fast-variable parameter. The symbols $r$ and $x_R$
denote the adaptation rate and reference level. The equations are
\begin{align}
\dot x &= y-x^3+\beta x^2-z+I+c_1, \label{eq:hr1}\\
\dot y &= 1-5x^2-y, \label{eq:hr2}\\
\dot z &= r\,\big(4(x-x_R)-z\big)+c_s. \label{eq:hr3}
\end{align}
The reference parameter set is
\begin{equation}
(r,x_R,\beta,I)=(0.01,-1.6,2.90,3.60).
\label{eq:hrpar}
\end{equation}
The model has no forcing. The variable $z$ is its slowest recovery variable.

Each intervention modifies the fast equation or the slowest recovery equation.
Let $f_1(\bm x,t)$ denote the uncontrolled right-hand side of the fast
equation. \emph{Recovery damping} uses $c_s=-k x_s$ with gain $k\ge0$ and
$c_1=0$. \emph{Rate scaling} divides the fast equation by $1+\gamma$ with $\gamma\ge0$.
It changes the time-scale ratio between the fast and slow variables. It is a
parameter change, not a feedback law. Its effort is measured through the
equivalent additive input $c_1=-\gamma f_1/(1+\gamma)$.

\emph{Tracking} uses $c_1=-\kappa\,\big(x_1-\bar x_1(t)\big)$ with gain
$\kappa\ge0$ and $c_s=0$. Here $\bar{\bm x}(t)$ is an unstable periodic orbit of
the uncontrolled model, found by Newton iteration on a return map. The return
map is the stroboscopic map for the forced models. For the Hindmarsh--Rose
model it is the map between successive maxima of $z$. The law is that of \cite{chendong1993}. That work approximated the orbit by harmonic balance instead.

For Pyragas control, let $j$ denote the controlled component and let $\tau$
denote the delay. The choice is $j=s$ in the FitzHugh--Nagumo model and $j=1$
in the other two models. \emph{Pyragas control} uses
$c_j=\eta\,\big(x_j(t-\tau)-x_j(t)\big)$ with gain $\eta\ge0$
\cite{pyragas}. The delay is $\tau=T$ in the forced models. In the
Hindmarsh--Rose model, $\tau=T_u$, where $T_u$ is the period of the target
orbit.

The target orbits are the following. The FitzHugh--Nagumo target is a period-1
orbit that fires once per forcing period. The Aliev--Panfilov target is a
period-2 orbit that fires once every two forcing periods. The Hindmarsh--Rose
target has one spike per period $T_u$. Each target lies inside the chaotic
attractor of its model.

Chaos is measured by the largest state Lyapunov exponent. For the ordinary
differential equation cases, perturbations obey
$\dot{\delta\bm x}=A_c(t)\,\delta\bm x$, where $A_c(t)$ is the closed-loop
Jacobian along the trajectory. Under Pyragas control, the tangent equation also
contains the delayed perturbation and evolves on the corresponding history
space. Let $\mathcal U(t,0)$ denote the tangent evolution operator. It reduces
to the fundamental matrix in the ordinary differential equation cases. The
largest state Lyapunov exponent is
\begin{equation}
\lambda_1=\limsup_{t\to\infty}\frac1t\log\|\mathcal U(t,0)\|,
\label{eq:lle}
\end{equation}
where $\|\cdot\|$ is a compatible operator norm. It is computed with the
Benettin algorithm \cite{benettin}. In the forced models, this exponent is
conditional on the prescribed forcing phase, and regularity means
$\lambda_1<0$. In the autonomous Hindmarsh--Rose model, it is the ordinary
state exponent. A periodic orbit has $\lambda_1=0$ along the direction of the
flow unless a time-varying input breaks the time-shift symmetry. Define
\begin{equation}
\lambda_\epsilon=10^{-3}.
\label{eq:tol}
\end{equation}
Regularity in the autonomous model means $\lambda_1\le\lambda_\epsilon$.
The tolerance is about six percent of the uncontrolled exponent.

The firing threshold $\theta$ is the middle zero of the fast nonlinearity at
zero recovery and zero drive. It gives $\theta=0$ for the FitzHugh--Nagumo and
Hindmarsh--Rose models and $\theta=\alpha$ for the Aliev--Panfilov model. It is
a reproducible classification rule, not a biophysical constant. Let $T_w$ be a
reference window, equal to $T$ in the forced models and to $T_u$ in the
Hindmarsh--Rose model. Let $N_w$ be the number of observed windows. For each
window $n=1,\ldots,N_w$, define
\begin{equation}
\chi_n=
\begin{cases}
1, & x_1 \text{ crosses } \theta \text{ upward during window } n,\\
0, & \text{otherwise}.
\end{cases}
\label{eq:chi}
\end{equation}
The excitation index is
\begin{equation}
E=\frac{1}{N_w}\sum_{n=1}^{N_w}\chi_n.
\label{eq:E}
\end{equation}
A cell held above threshold has $E=0$. Let $x_1^{0}(t)$ denote the uncontrolled
fast variable at the same drive. The excursion ratio is
\begin{equation}
R=\frac{\max_t x_1-\min_t x_1}{\max_t x_1^{0}-\min_t x_1^{0}}.
\label{eq:R}
\end{equation}
Firing with $R\ge 1/2$ is called substantial.
On an observation interval $[t_a,t_b]$, let $N_\uparrow$ and $N_\uparrow^0$
denote the upward crossing counts for the controlled and uncontrolled systems.
The firing-rate ratio is
\begin{equation}
S=\frac{N_\uparrow}{N_\uparrow^0}.
\label{eq:S}
\end{equation}
Let $c(t)$ denote the active additive control input. For rate scaling, $c(t)$ is
the equivalent input $c_1(t)$ defined above. The root-mean-square effort is
\begin{equation}
Q=\left[\frac{1}{t_b-t_a}\int_{t_a}^{t_b}c^2(t)\,dt\right]^{1/2}.
\label{eq:Q}
\end{equation}

The exponent can be nonmonotone in gain. When
a scan ends in one contiguous regular interval, $g_{\rm on}$ denotes the lower
edge of that interval. Aliev--Panfilov Pyragas control has no such interval.
For that case $g_{\rm on}$ denotes the lower edge of the principal low-gain
regular band, which ends at $\eta=1.60$ and is followed by alternating
narrow bands. The silencing gain $g_s$ is the
smallest scanned gain from which every larger scanned gain gives $E=0$.

All runs use fourth-order Runge--Kutta. Let $h$
denote the integration step. The values are
\begin{equation}
h=\begin{cases}
T/300, & \text{FitzHugh--Nagumo},\\
T/350, & \text{Aliev--Panfilov},\\
\text{approximately }0.05, & \text{Hindmarsh--Rose}.
\end{cases}
\label{eq:step}
\end{equation}
For the Hindmarsh--Rose model, $h$ is adjusted to give an integer number of
steps per $T_u$. Transients are discarded. Let $T_o$ denote the observation
length. The values are
\begin{equation}
T_o=\begin{cases}
200T, & \text{forced models},\\
30\,000, & \text{Hindmarsh--Rose}.
\end{cases}
\label{eq:obs}
\end{equation}
Tracking targets are sampled at half steps. The Pyragas history is interpolated
linearly at half steps. The threshold gains come from one-dimensional scans with
fine gain steps.

\section{Chaotic Regimes and Silencing over Gain and Drive}\label{sec:maps}

\begin{figure}[htbp]
\centering
\begin{subfigure}[b]{0.32\linewidth}
  \centering
  \includegraphics[width=\linewidth]{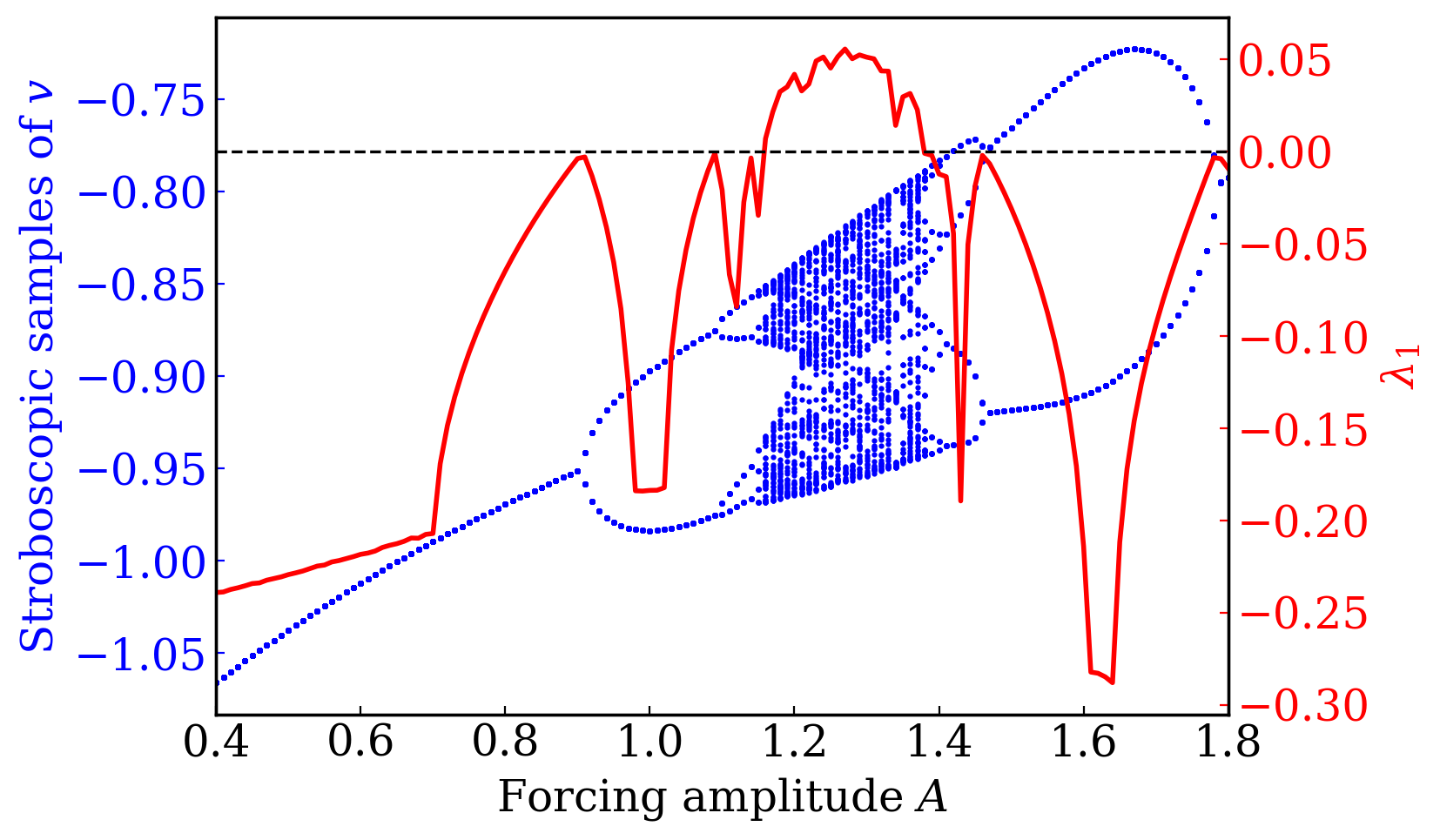}
  \caption{}
\end{subfigure}
\hfill
\begin{subfigure}[b]{0.32\linewidth}
  \centering
  \includegraphics[width=\linewidth]{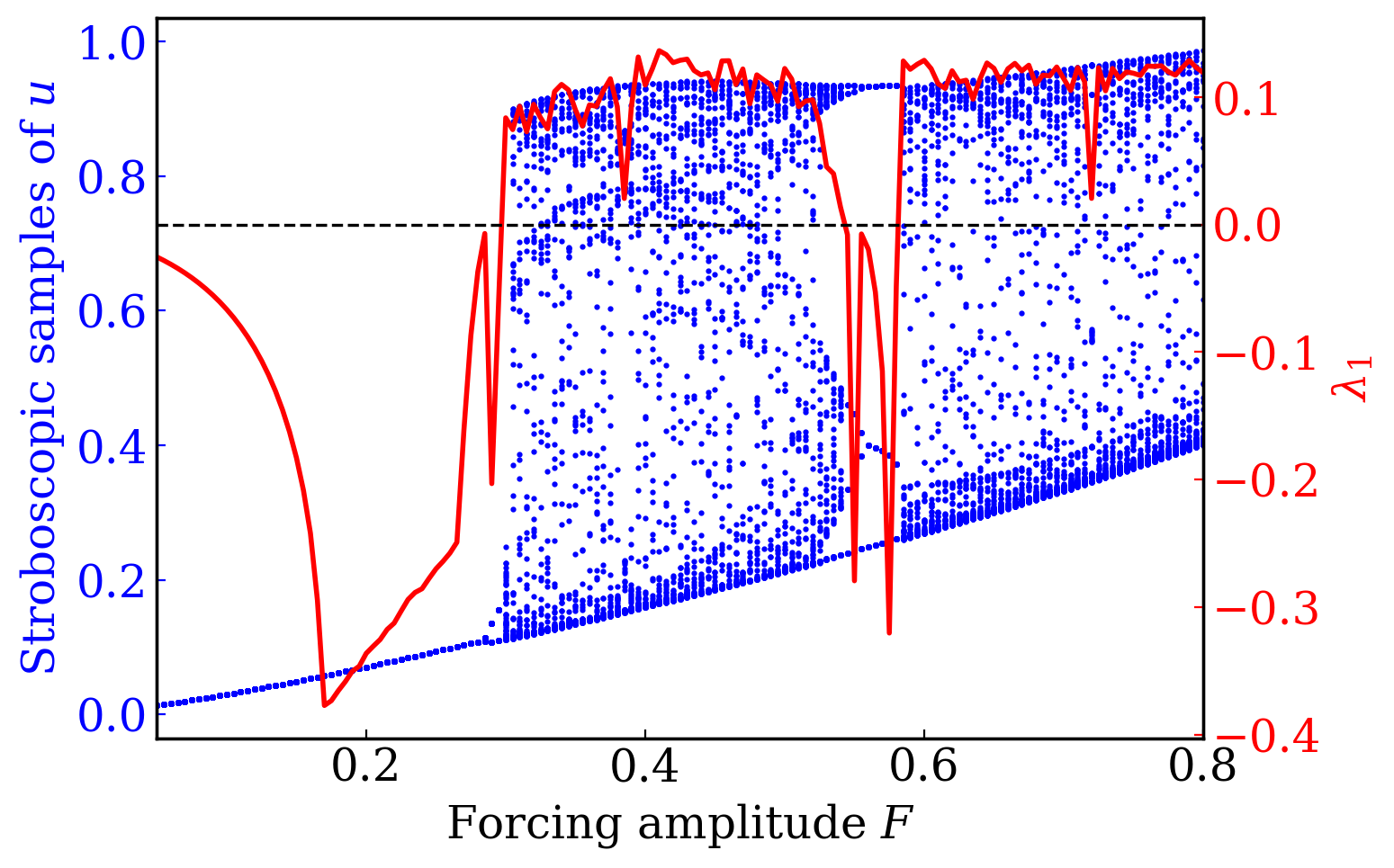}
  \caption{}
\end{subfigure}
\hfill
\begin{subfigure}[b]{0.32\linewidth}
  \centering
  \includegraphics[width=\linewidth]{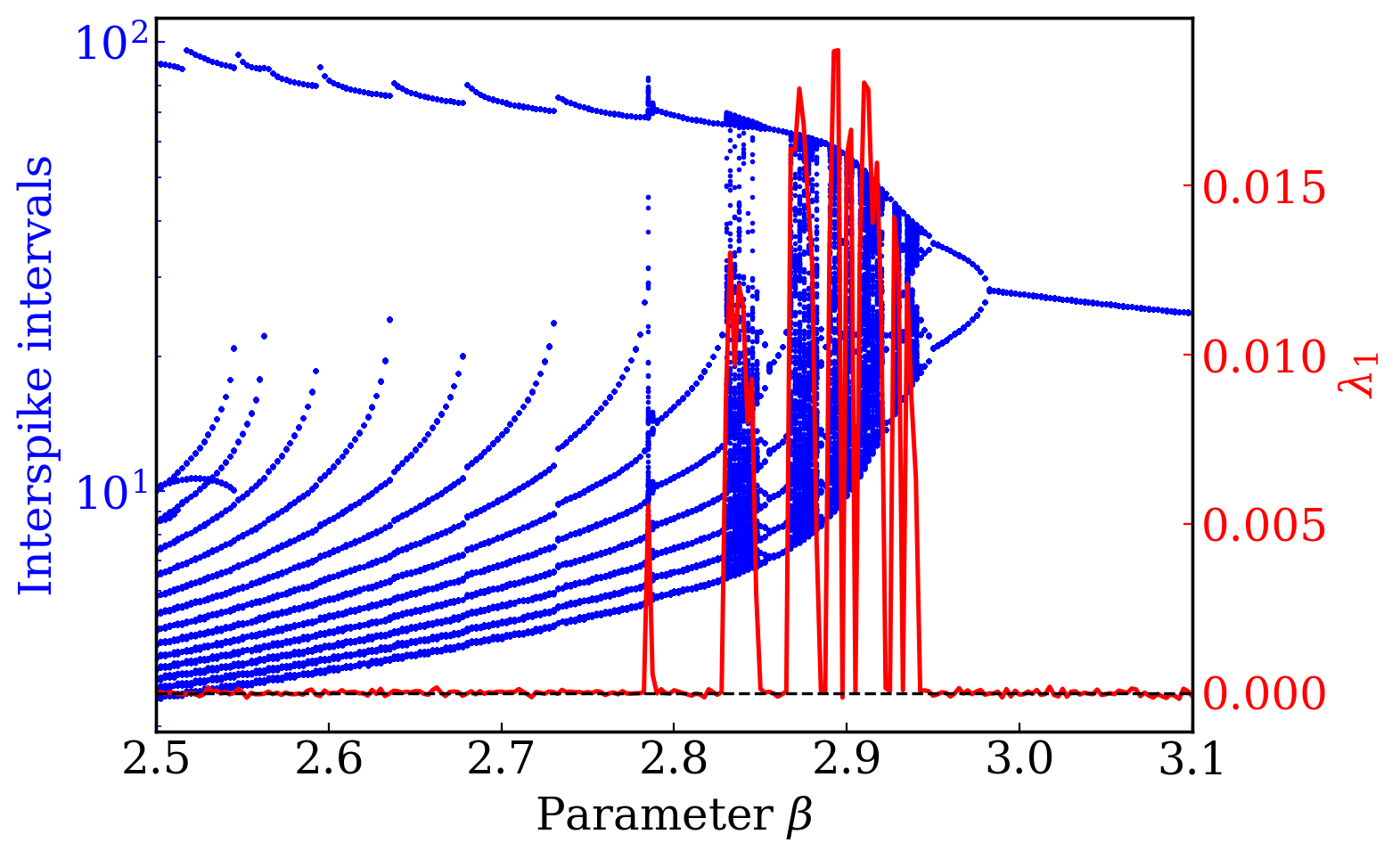}
  \caption{}
\end{subfigure}
\caption{Uncontrolled bifurcation diagrams with the state Lyapunov
exponent overlaid. Blue dots are samples of the fast dynamics. The red curve is
$\lambda_1$, and the dashed line marks $\lambda_1{=}0$. (a)~FitzHugh--Nagumo
model, stroboscopic samples of $v$ against forcing amplitude $A$.
(b)~Aliev--Panfilov model, stroboscopic samples of $u$ against forcing
amplitude $F$. (c)~Hindmarsh--Rose model at $I{=}3.60$, interspike intervals
against $\beta$, on a logarithmic scale.}
\label{fig:bif}
\end{figure}

Fig.~\ref{fig:bif}(a) shows the uncontrolled FitzHugh--Nagumo model. A
stroboscopic branch period-doubles near $A\approx0.9$, and the exponent becomes
positive in a band around $A=1.3$. A reverse sequence closes the band. Nearby
periodic responses can still look complicated in the phase plane. The sign of
$\lambda_1$ is therefore used together with the stroboscopic section.

Fig.~\ref{fig:bif}(b) shows the uncontrolled Aliev--Panfilov model. The response
is period-1 at small forcing. Higher-period structure follows, and the exponent
becomes positive near $F\approx0.30$. A periodic window with a negative exponent
interrupts the chaotic band near $F\approx0.55$. The reference value $F=0.40$
lies inside the chaotic region.

Fig.~\ref{fig:bif}(c) shows the uncontrolled Hindmarsh--Rose model at $I=3.60$
as the parameter $\beta$ varies. The interspike intervals form ordered branches
of periodic bursting at small $\beta$ and a single tonic branch at large
$\beta$. The exponent is positive only in a narrow band near $\beta=2.9$, which
contains periodic windows. The reference value $\beta=2.90$ lies inside that
band. Chaos at this operating point is therefore fragile. A small change of
$\beta$ alone ends it.

\begin{figure}[htbp]
\centering
\begin{subfigure}[b]{0.49\linewidth}
  \centering
  \includegraphics[width=\linewidth]{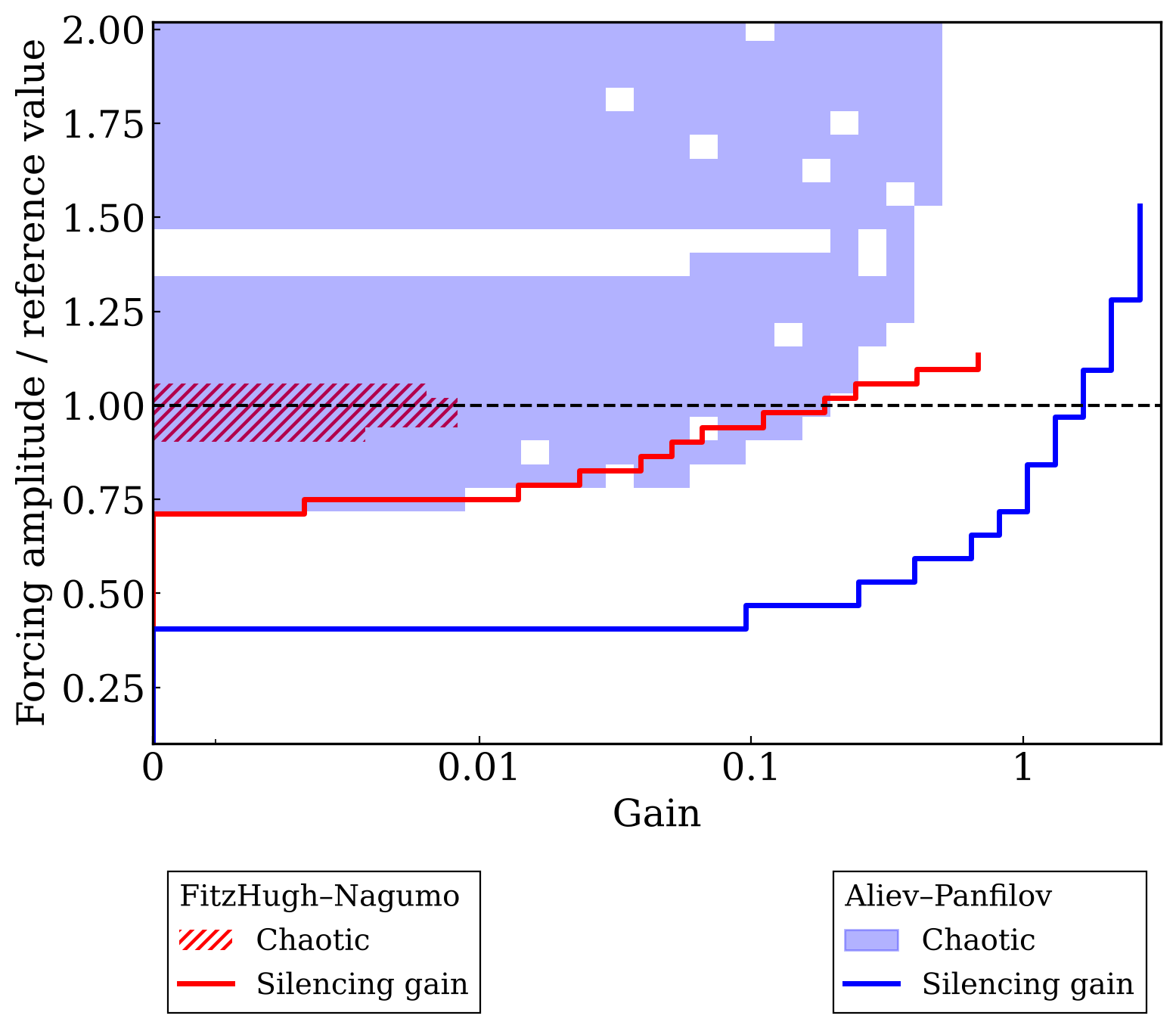}
  \caption{}
\end{subfigure}
\hfill
\begin{subfigure}[b]{0.49\linewidth}
  \centering
  \includegraphics[width=\linewidth]{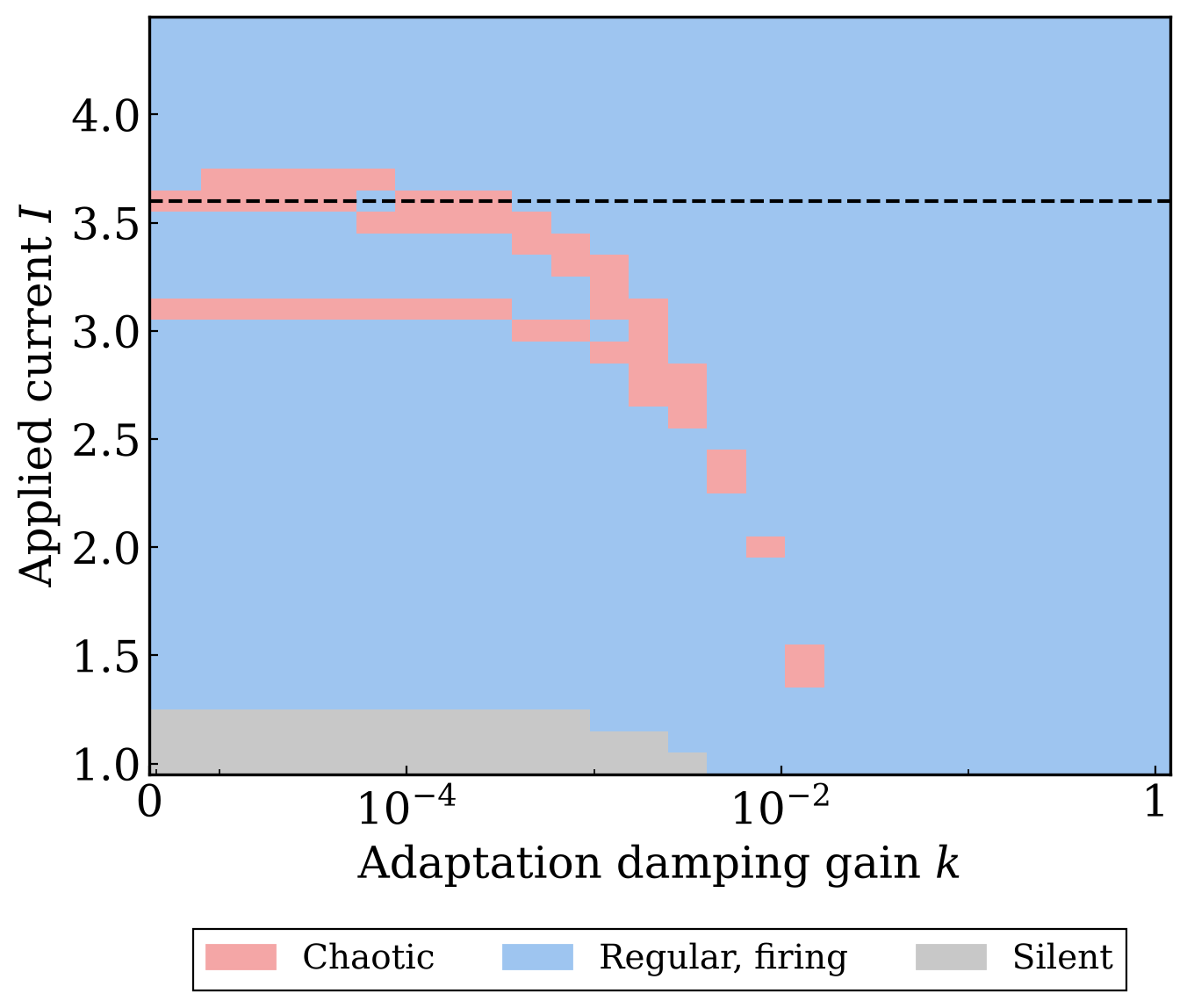}
  \caption{}
\end{subfigure}
\caption{Chaotic, firing, and silent regions over gain and drive.
(a)~Forced models. The gain is recovery damping $k$ for the FitzHugh--Nagumo
model and rate scaling $\gamma$ for the Aliev--Panfilov model. The forcing
amplitude is divided by its reference value. Chaotic cells have
$\lambda_1{>}0$. (b)~Hindmarsh--Rose model under recovery damping of $z$.
Chaotic cells have $\lambda_1{>}\lambda_\epsilon$. Dashed lines mark the
reference regimes. Gain axes are linear near zero and logarithmic above.}
\label{fig:maps}
\end{figure}

Fig.~\ref{fig:maps}(a) maps the two forced models in one panel. It shows
recovery damping in the FitzHugh--Nagumo model and rate scaling in the
Aliev--Panfilov model. The horizontal axis is the gain, $k$ or $\gamma$. The
vertical axis is the forcing amplitude divided by its reference value, so the
dashed line marks both reference regimes. Hatched red and shaded blue cells are
chaotic. To the right of each solid line, the cell is silent. Between the
chaotic cells and the solid line of the same color, the cell fires under the
stated threshold. The excursion ratio decides whether that firing is
substantial. In the FitzHugh--Nagumo model, chaos occupies a narrow band of
forcing and ends below a gain of 0.01. Firing persists to a gain more than an
order of magnitude larger. In the Aliev--Panfilov model, chaos covers a wide
range of forcing, and the silencing line lies far to the right. Both silencing
gains rise with the forcing amplitude.

Fig.~\ref{fig:maps}(b) maps the Hindmarsh--Rose model under recovery damping of
the adaptation variable, with the applied current $I$ as the drive. The
picture differs from the forced models. The silent region lies at small
current and small gain, and damping shrinks it. Chaotic cells form a narrow
band that moves to lower current as the gain grows. No gain silences the cell
at the reference current.

Table~\ref{tab:thresholds} lists the gain intervals at the reference regimes for
all four interventions. In the FitzHugh--Nagumo model, the silencing gain under
recovery damping lies more than an order of magnitude above the onset of the
regular band. In the Aliev--Panfilov model, recovery damping removes chaos and
firing at the same scanned gain, and the cell then stays depolarized near
$u\approx1$. Under rate scaling in that model, chaos first disappears near
$\gamma=0.19$, returns in a narrow band near $\gamma=0.24$, and is absent at
every larger scanned gain. In the Hindmarsh--Rose model, no intervention
silences the cell in the scanned range. Tracking and Pyragas control silence no
model.

\begin{table}[htbp]
\vspace{14pt}
\renewcommand{\arraystretch}{1.15}
\tbl{Gain thresholds at the reference regimes. $g_{\rm on}$: onset of the
terminal regular interval. $g_s$: silencing gain. Each value is the upper end
of an interval between adjacent scanned gains. A dash means that no scanned
gain silenced the cell. No intervention silenced the Hindmarsh--Rose cell, so
its $g_s$ row is omitted. The value in parentheses uses the stricter threshold
$u{=}0.5$.\label{tab:thresholds}}
{\begin{tabular}{|l|c|c|c|c|}
\hline
 & Damping $k$ & Rate $\gamma$ & Tracking $\kappa$ & Pyragas $\eta$ \\
\hline
\multicolumn{5}{|l|}{FitzHugh--Nagumo} \\
\hline
$g_{\rm on}$ & 0.009 & 0.015 & 0.012 & 0.015 \\
\hline
$g_s$ & 0.175 & 0.225 & -- & -- \\
\hline
\multicolumn{5}{|l|}{Aliev--Panfilov} \\
\hline
$g_{\rm on}$ & 0.1825 & 0.25 & 0.32 & 0.22 \\
\hline
$g_s$ & 0.1825 & 1.60 (0.27) & -- & -- \\
\hline
\multicolumn{5}{|l|}{Hindmarsh--Rose} \\
\hline
$g_{\rm on}$ & 0.000375 & 0.1425 & 1.10 & 0.39 \\
\hline
\end{tabular}}
\end{table}

\section{Four Interventions on the Same Chaotic Regimes}\label{sec:compare}

\begin{figure}[htbp]
\centering
\begin{subfigure}[b]{0.32\linewidth}
  \centering
  \includegraphics[width=\linewidth]{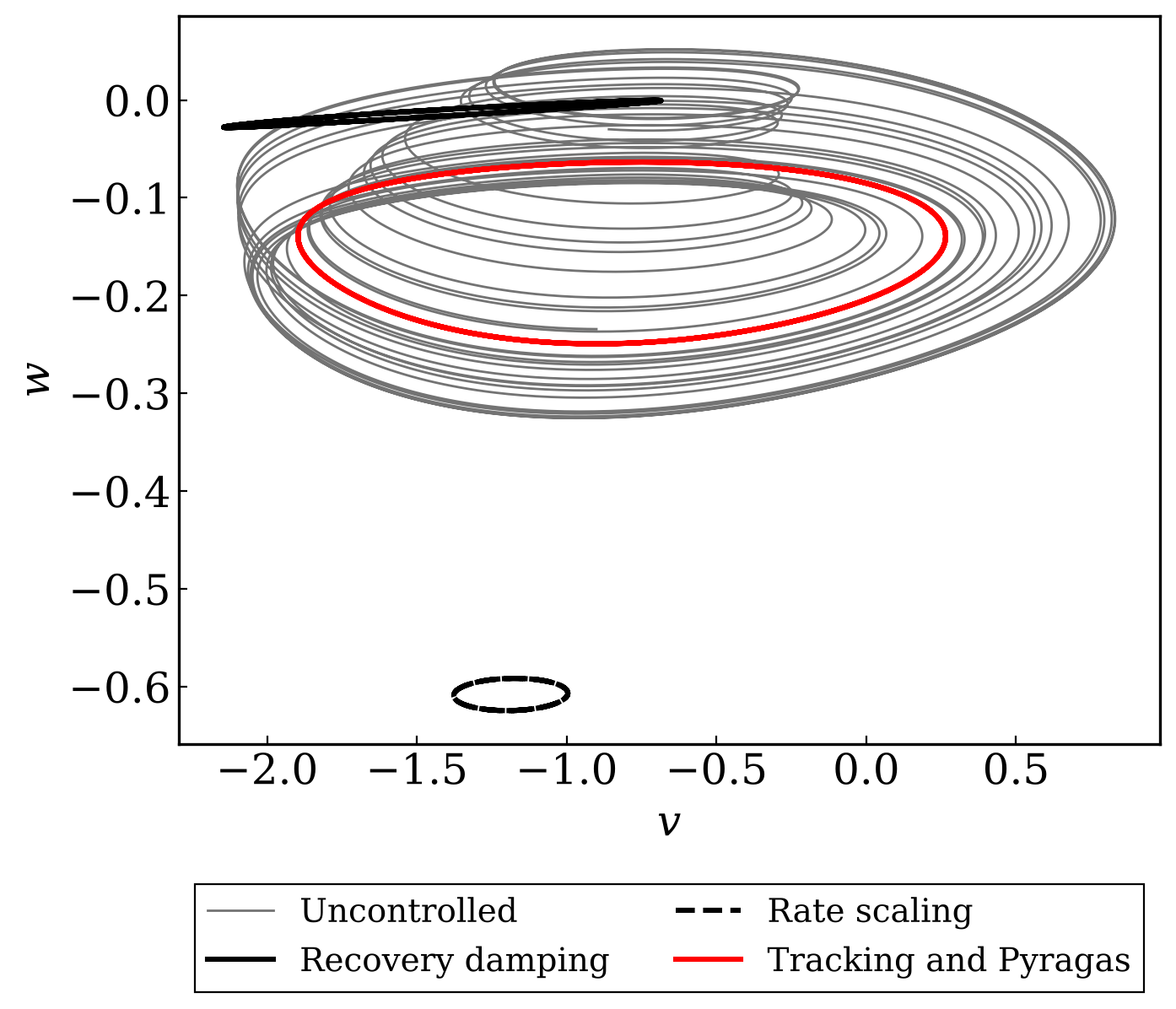}
  \caption{}
\end{subfigure}
\hfill
\begin{subfigure}[b]{0.32\linewidth}
  \centering
  \includegraphics[width=\linewidth]{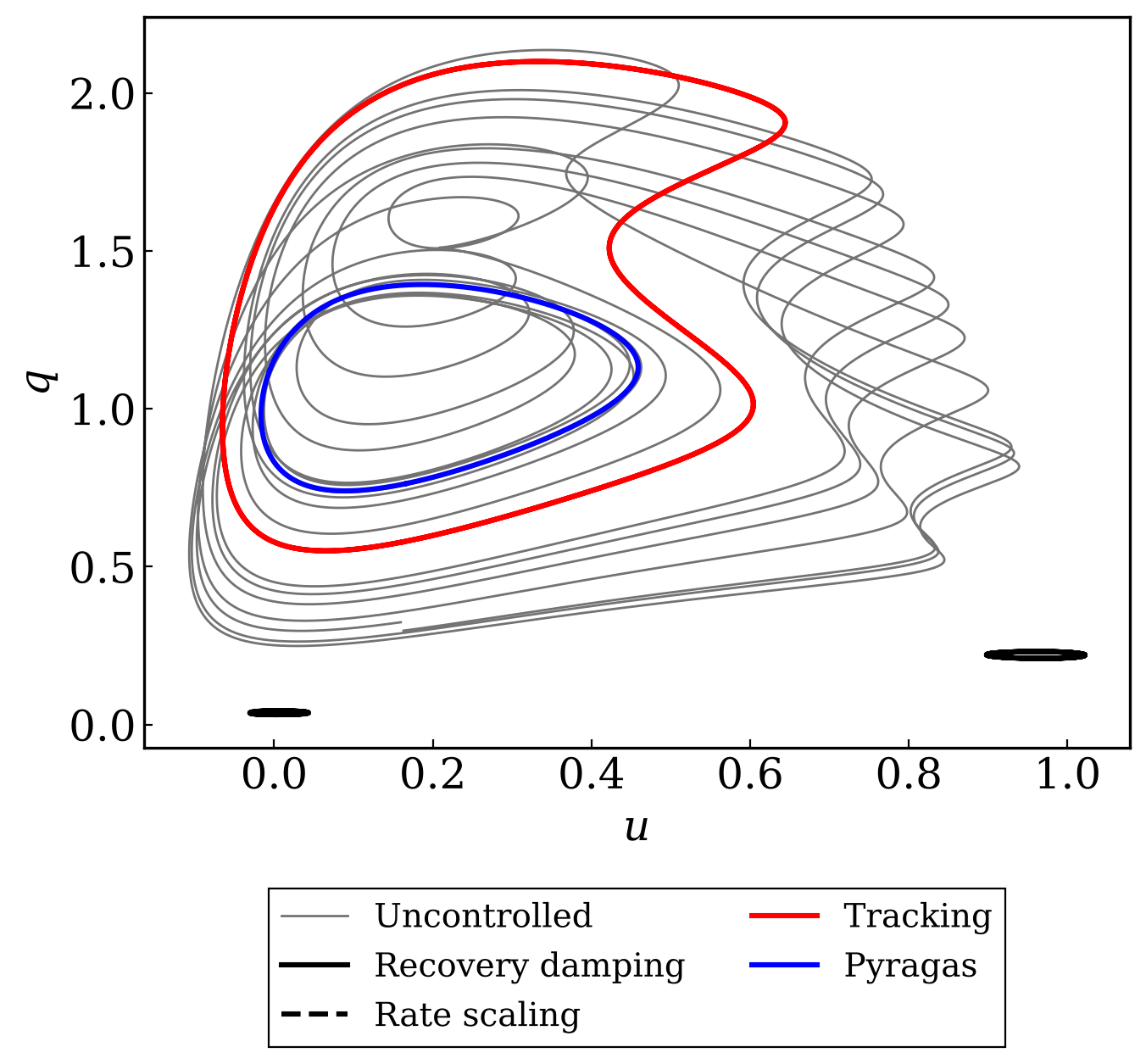}
  \caption{}
\end{subfigure}
\hfill
\begin{subfigure}[b]{0.32\linewidth}
  \centering
  \includegraphics[width=\linewidth]{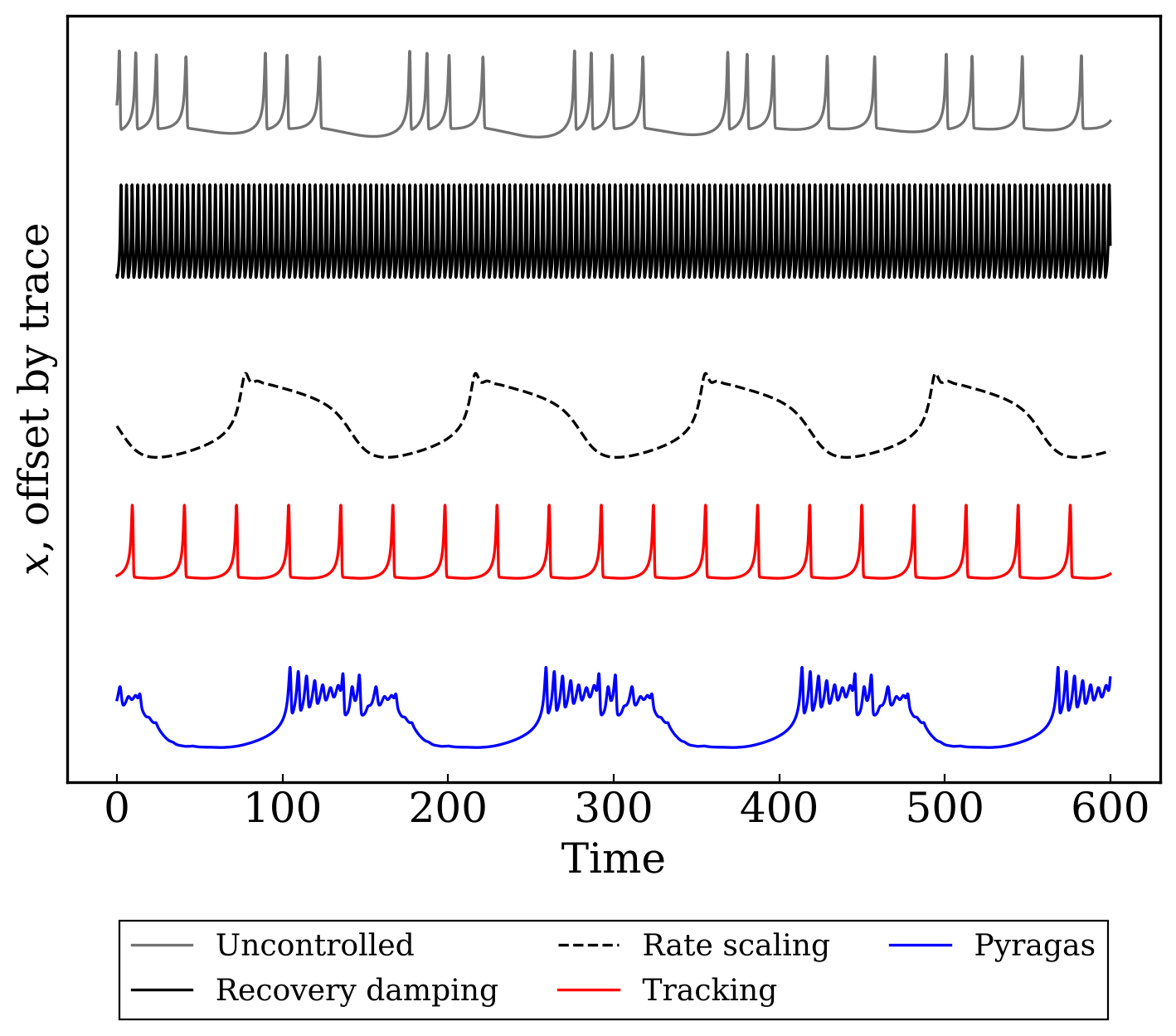}
  \caption{}
\end{subfigure}
\caption{Controlled responses at the reference regimes. Gray: uncontrolled.
Black solid: recovery damping. Black dashed: rate scaling. Red: tracking.
Blue: Pyragas control. (a)~FitzHugh--Nagumo orbits with $k{=}5$, $\gamma{=}5$,
and $\kappa{=}1$. Pyragas control with $\eta{=}0.2$ reaches the red orbit.
(b)~Aliev--Panfilov orbits with $k{=}0.2$, $\gamma{=}5$, $\kappa{=}1.5$, and
$\eta{=}0.4$. (c)~Hindmarsh--Rose time series of $x$, offset for clarity, with
$k{=}0.1$, $\gamma{=}5$, $\kappa{=}1.5$, and $\eta{=}0.5$.}
\label{fig:orbits}
\end{figure}

Fig.~\ref{fig:orbits}(a) shows the attracting orbits of the FitzHugh--Nagumo
model at the reference forcing. The gray curve is the chaotic attractor. Large
recovery damping pins $w$ near zero and keeps $v$ on the resting branch. Large
rate scaling contracts the orbit to a small loop below threshold. Tracking and
Pyragas control both converge to the same period-1 orbit, which lies inside the
chaotic attractor and crosses threshold every period.

Fig.~\ref{fig:orbits}(b) shows the Aliev--Panfilov model. Recovery damping holds
the cell in a depolarized state near $u\approx1$. Large rate scaling leaves a
small loop near the resting state. Tracking converges to the period-2 target.
Pyragas control converges to a period-1 orbit. Both orbits are embedded in the
chaotic attractor and cross threshold.

Fig.~\ref{fig:orbits}(c) shows time series of the Hindmarsh--Rose fast variable.
The uncontrolled cell fires irregular bursts. Recovery damping turns it into a
dense tonic spike train. Large rate scaling leaves slow waves whose peaks cross threshold only occasionally.
Tracking produces one regular spike per period of the target orbit. Pyragas
control produces regular bursts, but not the target orbit.

\begin{figure}[htbp]
\centering
\begin{subfigure}[b]{0.48\linewidth}
  \centering
  \includegraphics[width=0.92\linewidth]{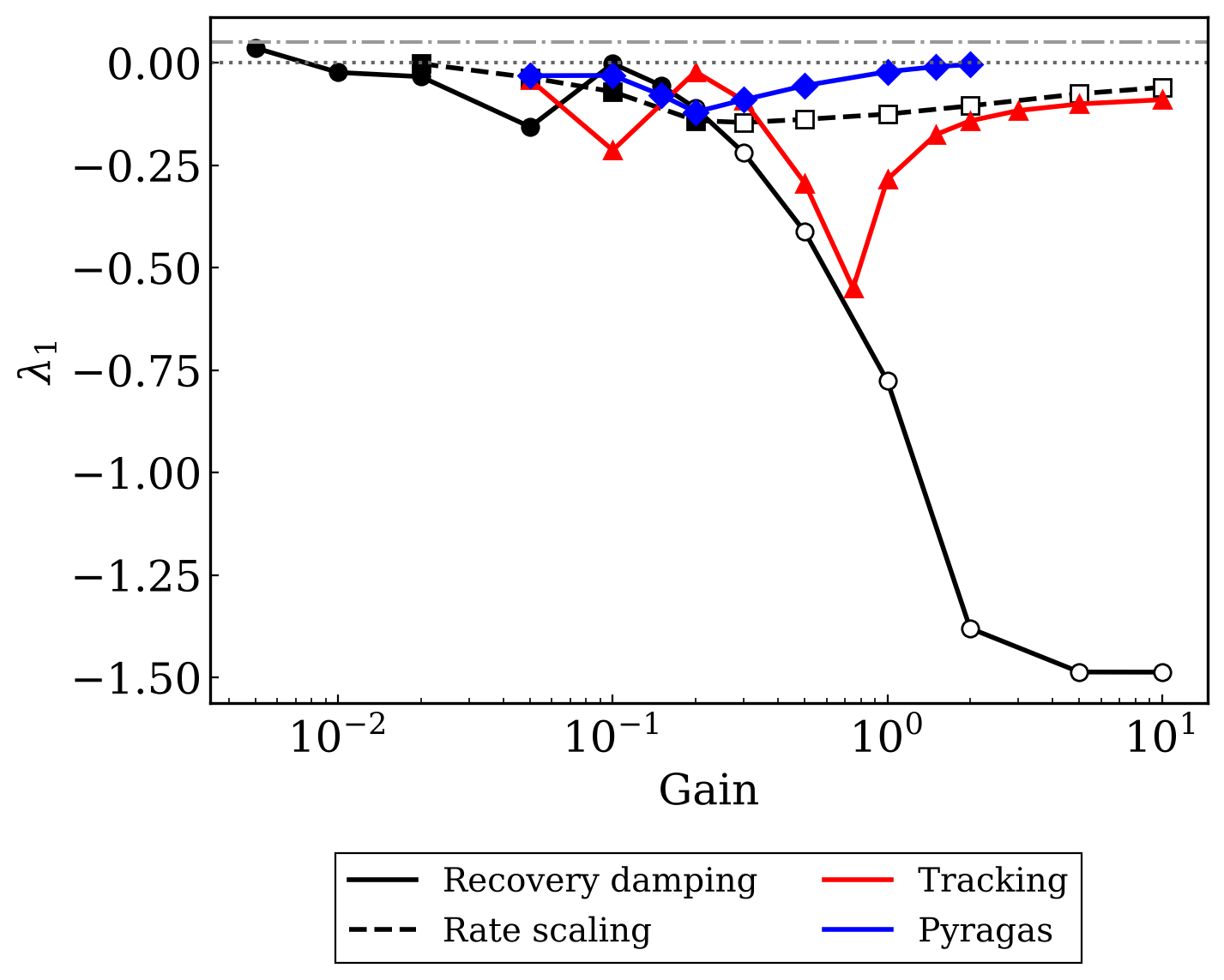}
  \caption{}
\end{subfigure}
\hfill
\begin{subfigure}[b]{0.46\linewidth}
  \centering
  \includegraphics[width=0.92\linewidth]{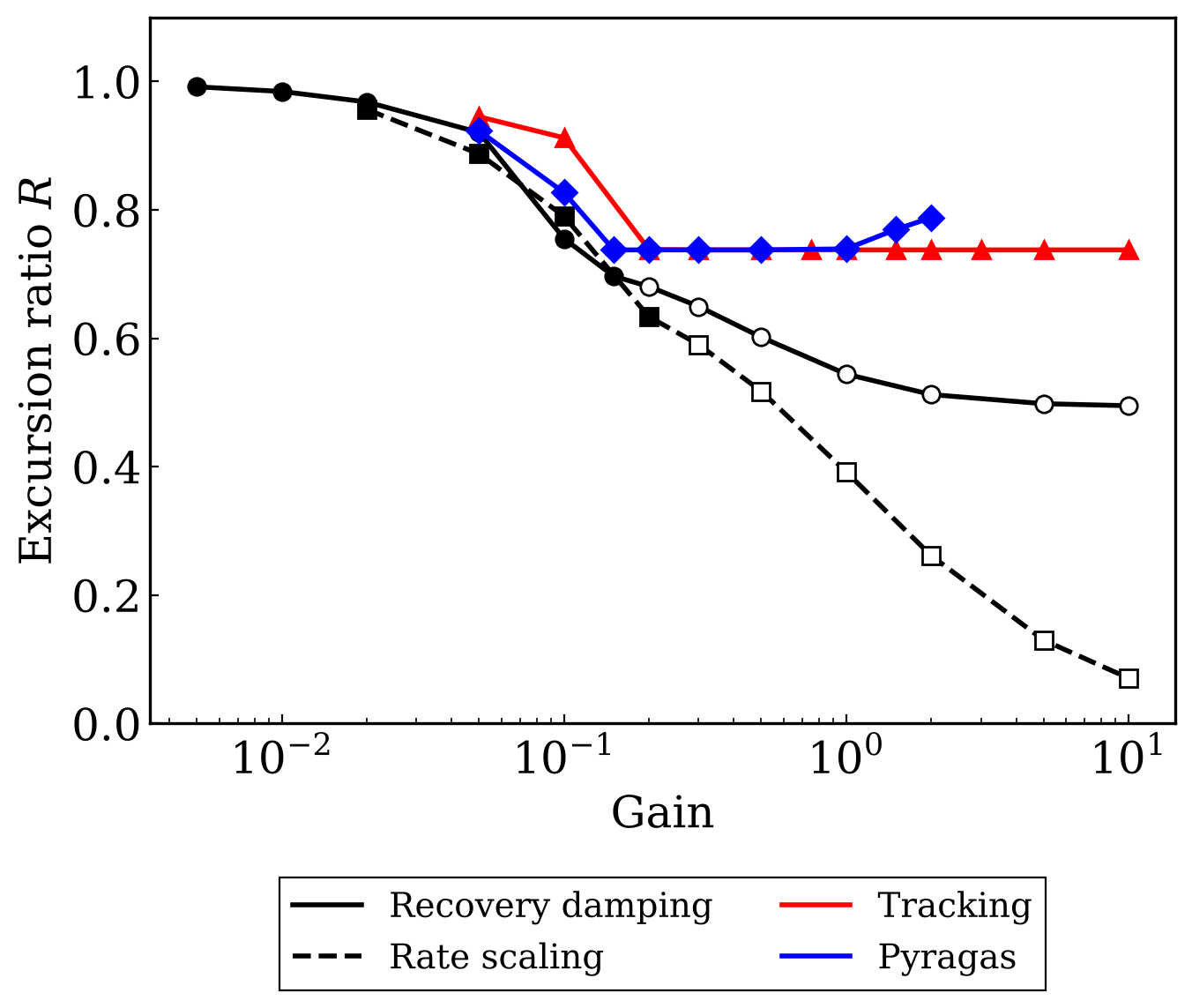}
  \caption{}
\end{subfigure}

\begin{subfigure}[b]{0.48\linewidth}
  \centering
  \includegraphics[width=0.92\linewidth]{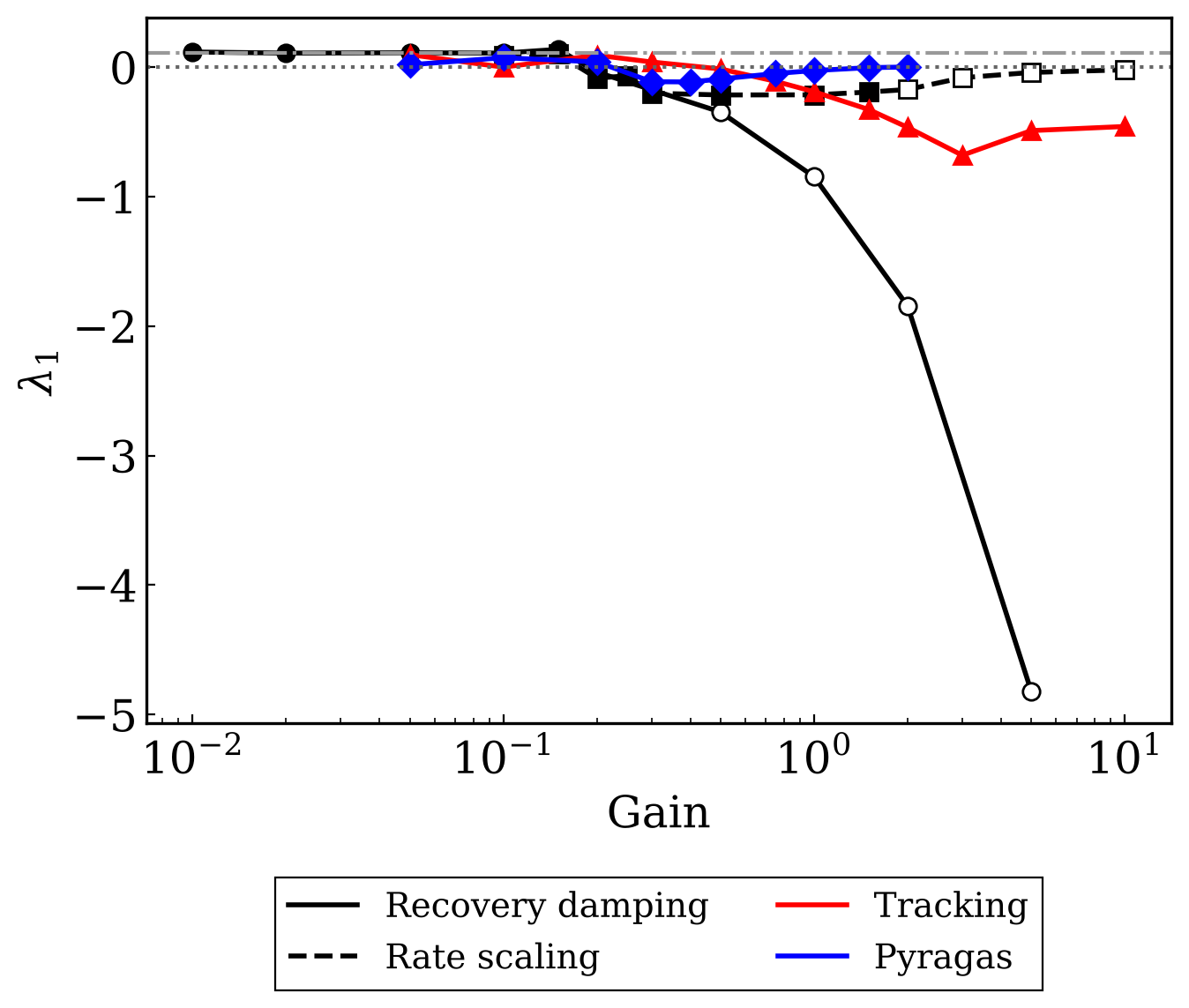}
  \caption{}
\end{subfigure}
\hfill
\begin{subfigure}[b]{0.48\linewidth}
  \centering
  \includegraphics[width=0.92\linewidth]{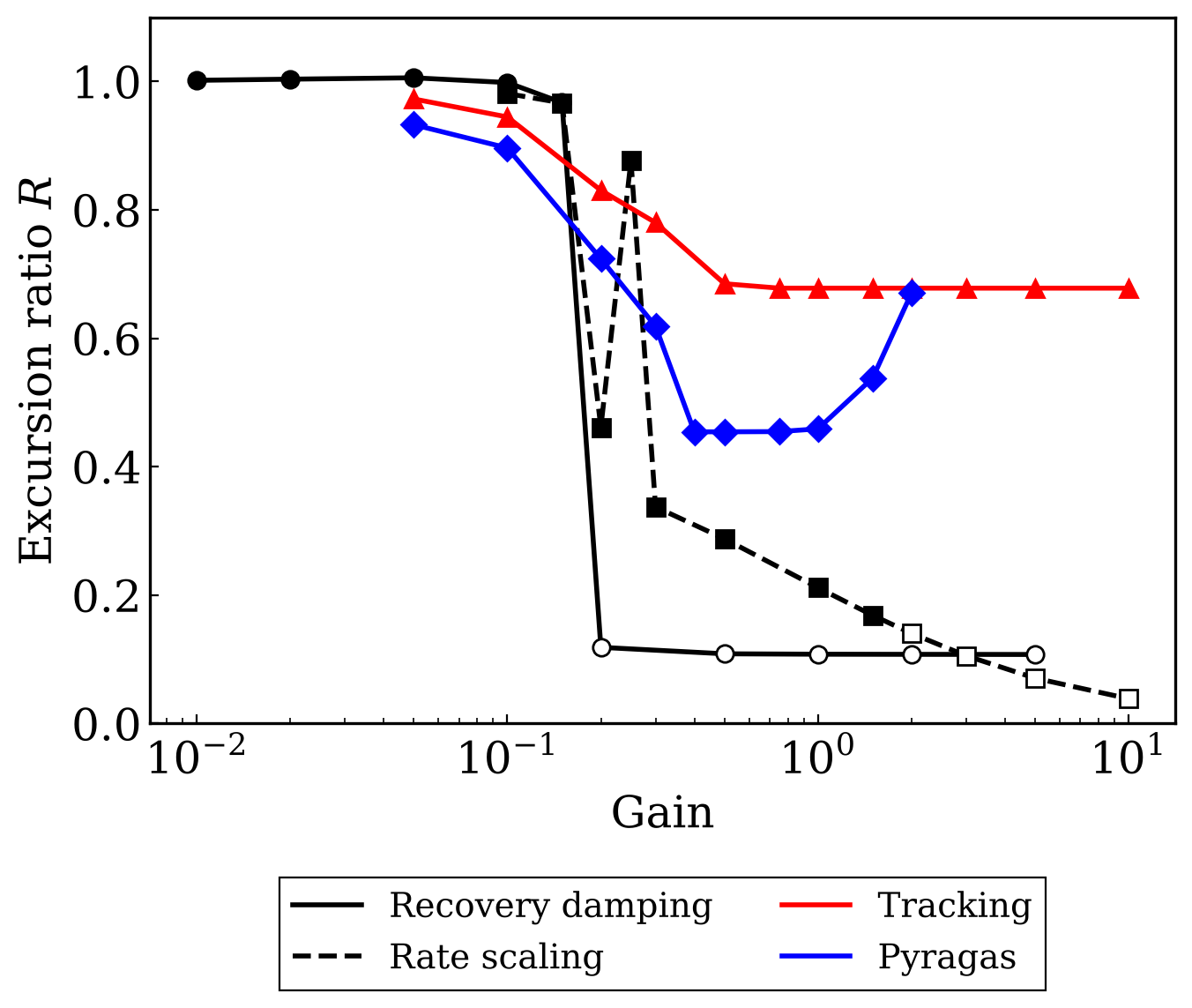}
  \caption{}
\end{subfigure}

\begin{subfigure}[b]{0.48\linewidth}
  \centering
  \includegraphics[width=0.92\linewidth]{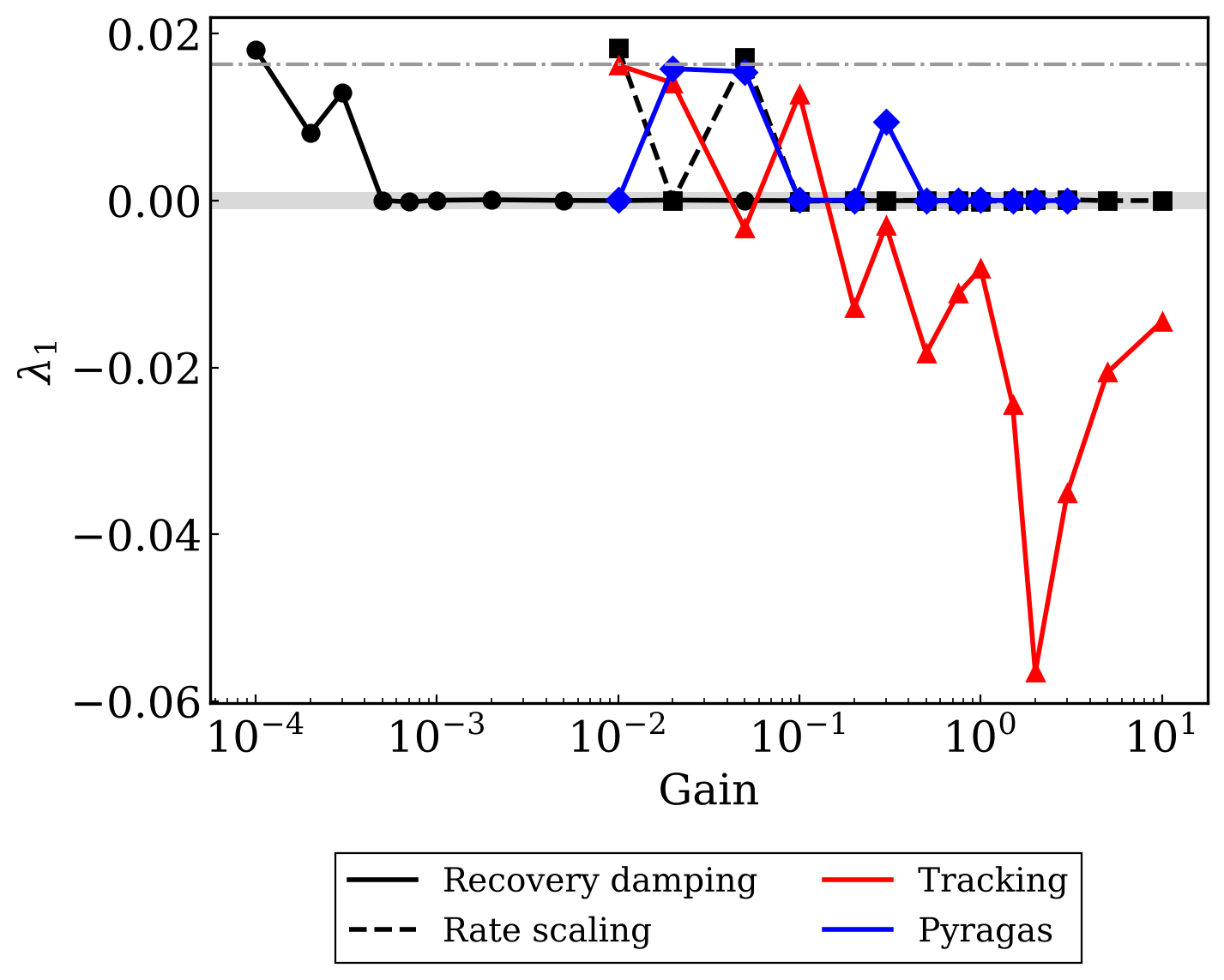}
  \caption{}
\end{subfigure}
\hfill
\begin{subfigure}[b]{0.48\linewidth}
  \centering
  \includegraphics[width=0.92\linewidth]{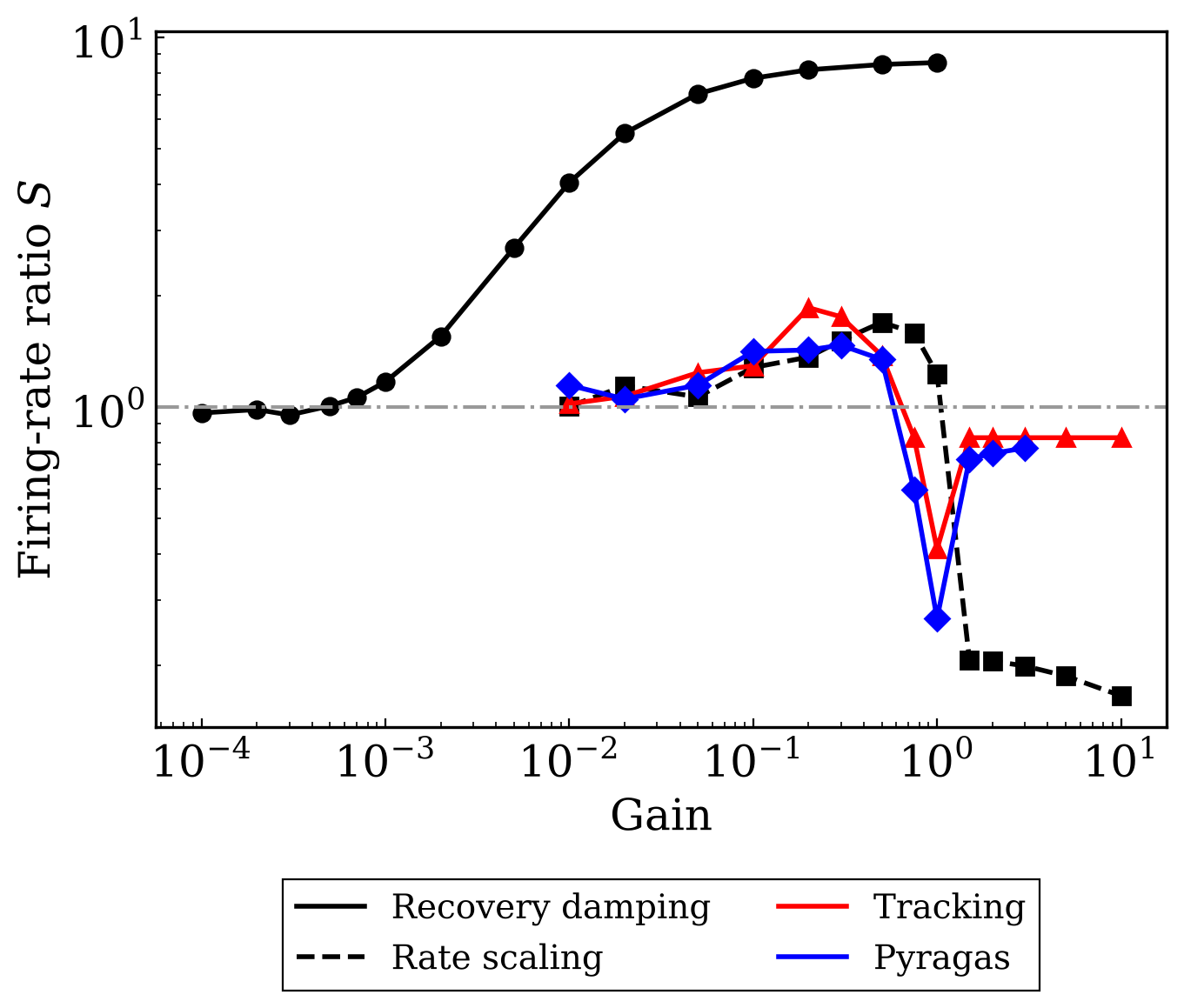}
  \caption{}
\end{subfigure}
\caption{Gain sweeps at the reference regimes. Filled markers: firing, $E{>}0$.
Open markers: silent, $E{=}0$. In (a), (c), (e), and (f), the dash-dotted line
marks the uncontrolled value. (a)~FitzHugh--Nagumo exponent $\lambda_1$. (b)~FitzHugh--Nagumo
excursion ratio $R$. (c)~Aliev--Panfilov exponent $\lambda_1$.
(d)~Aliev--Panfilov excursion ratio $R$. (e)~Hindmarsh--Rose exponent
$\lambda_1$, with the gray band marking $|\lambda_1|{\le}\lambda_\epsilon$.
(f)~Hindmarsh--Rose firing-rate ratio $S$, on a logarithmic scale.}
\label{fig:sweeps}
\end{figure}

Fig.~\ref{fig:sweeps}(a) shows $\lambda_1$ against gain for the four
interventions in the FitzHugh--Nagumo model. Filled markers denote firing and
open markers denote silence. Each intervention has a regular band, and the
exponent is not monotone in gain for any of them. Recovery damping reaches the
most negative exponents, but only in the silent range.
Fig.~\ref{fig:sweeps}(b) shows the excursion ratio. Damping and rate scaling
shrink the excursion steadily, and the markers turn open. Over most of their
gain range, tracking and Pyragas control hold the excursion at the value of the
target orbit.

Fig.~\ref{fig:sweeps}(c) shows the exponent for the Aliev--Panfilov model.
Recovery damping gives large negative exponents only after the cell
depolarizes. Rate scaling gives regular responses from $\gamma\approx0.25$.
Pyragas control gives a regular band that breaks into narrow pieces above
$\eta\approx1.6$. Fig.~\ref{fig:sweeps}(d) shows the excursion ratio. Under rate
scaling, the excursion falls below one fifth of the uncontrolled value before
the cell falls silent.

Fig.~\ref{fig:sweeps}(e) shows the exponent for the Hindmarsh--Rose model. The
gray band marks the regularity tolerance. Recovery damping, rate scaling, and
Pyragas control give $\lambda_1$ within the tolerance once they regularize the
cell, as expected for a periodic orbit of an autonomous flow. Tracking gives
negative exponents, because its reference signal makes the closed loop
time-varying. Fig.~\ref{fig:sweeps}(f) shows the firing-rate ratio. Recovery
damping raises the firing rate several-fold. Rate scaling lowers it to about
one fifth at large gain. Tracking settles at the rate of the target orbit.
Fig.~\ref{fig:sweeps}(f) uses $S$ instead of $R$ because the spike amplitude
barely changes in this model.

Table~\ref{tab:compare} compares the interventions at representative gains.
Damping and rate scaling appear twice in some models: once inside the firing
window and once at a large gain. In the forced models, the large gains $k=5$
and $\gamma=5$ silence the cell. Tracking and Pyragas control keep firing. In
the forced models their steady effort $Q$ stays below $10^{-7}$, a level set by
the numerical accuracy of the target orbit. In the Hindmarsh--Rose model,
tracking also reaches its target, with $Q=6.0\times10^{-6}$. Pyragas control
there gives regular firing with $Q=0.48$, so it settles on an orbit other
than the target. The table omits the small Hindmarsh--Rose gains
$k=5\times10^{-4}$ and $\gamma=0.2$. At these gains the cell is already
regular and keeps firing. Recovery damping gives $E=1.00$, $R=0.85$, and
$S=1.01$. Rate scaling gives $E=0.68$, $R=1.15$, and $S=1.36$.

\begin{table}[htbp]
\vspace{14pt}
\renewcommand{\arraystretch}{1.2}
\tbl{Interventions at representative gains. $\lambda_1$: largest state
Lyapunov exponent. $E$: excitation index. $R$: excursion ratio. $S$:
firing-rate ratio. ${\approx}0$: within the regularity
tolerance.\label{tab:compare}}
{\begin{tabular}{|l|r|c|c|c|}
\hline
Intervention and gain & $\lambda_1$ & $E$ & $R$ & $S$ \\
\hline
\multicolumn{5}{|l|}{FitzHugh--Nagumo} \\
\hline
Uncontrolled & $+0.052$ & 0.60 & 1.00 & 1.00 \\
\hline
Damping, $k=0.05$ & $-0.156$ & 0.50 & 0.92 & 0.83 \\
\hline
Damping, $k=5$ & $-1.487$ & 0 & 0.50 & 0 \\
\hline
Rate, $\gamma=0.2$ & $-0.141$ & 1.00 & 0.63 & 1.65 \\
\hline
Rate, $\gamma=5$ & $-0.076$ & 0 & 0.13 & 0 \\
\hline
Tracking, $\kappa=1$ & $-0.284$ & 1.00 & 0.74 & 1.65 \\
\hline
Pyragas, $\eta=0.2$ & $-0.120$ & 1.00 & 0.74 & 1.65 \\
\hline
\multicolumn{5}{|l|}{Aliev--Panfilov} \\
\hline
Uncontrolled & $+0.107$ & 0.51 & 1.00 & 1.00 \\
\hline
Damping, $k=0.2$ & $-0.039$ & 0 & 0.12 & 0 \\
\hline
Rate, $\gamma=0.3$ & $-0.202$ & 1.00 & 0.34 & 1.98 \\
\hline
Rate, $\gamma=5$ & $-0.041$ & 0 & 0.07 & 0 \\
\hline
Tracking, $\kappa=1.5$ & $-0.328$ & 0.50 & 0.68 & 0.99 \\
\hline
Pyragas, $\eta=0.4$ & $-0.115$ & 1.00 & 0.45 & 1.98 \\
\hline
\multicolumn{5}{|l|}{Hindmarsh--Rose} \\
\hline
Uncontrolled & $+0.016$ & 0.79 & 1.00 & 1.00 \\
\hline
Damping, $k=0.1$ & ${\approx}0$ & 1.00 & 1.07 & 7.75 \\
\hline
Rate, $\gamma=5$ & ${\approx}0$ & 0.23 & 0.97 & 0.19 \\
\hline
Tracking, $\kappa=1.5$ & $-0.024$ & 1.00 & 0.84 & 0.83 \\
\hline
Pyragas, $\eta=0.5$ & ${\approx}0$ & 0.57 & 0.92 & 1.34 \\
\hline
\end{tabular}}
\end{table}

\section{Contraction Certificates and Excitation}\label{sec:cert}

This section asks whether a contraction certificate can prove regularity while
the cell still fires. For a constant matrix $P=P^\top\succ0$ and a real square
matrix $M$, define the $P$-weighted matrix measure
\begin{equation}
\mu_P(M)=\text{largest eigenvalue of }\tfrac12\big(B+B^\top\big),
\qquad B=P^{1/2}MP^{-1/2}.
\label{eq:muP}
\end{equation}
Let $\mathsf I$ denote the identity matrix. The identity metric
$P=\mathsf I$ gives the Euclidean measure $\mu_{\mathsf I}$. Along a
trajectory, define the time average
\begin{equation}
\bar\mu_P=\limsup_{t\to\infty}\frac1t\int_0^t\mu_P\big(A_c(\sigma)\big)\,d\sigma .
\label{eq:muPbar}
\end{equation}
The bound below is classical \cite{lozinskii1958,desoer1972,lohmiller1998}.

\begin{theorem}[Matrix-measure bound]\label{thm:bound}
(i)~For every constant $P\succ0$, the exponent \eqref{eq:lle} satisfies
$\lambda_1\le\bar\mu_P$. (ii)~Suppose $A_c=J-kD$, where $J$ is the Jacobian without the gain term, $k\ge0$ is the gain, and $D\succeq0$ is a constant symmetric matrix. Let $\hat\mu(k)$ be the supremum of
$\mu_{\mathsf I}\big(J(\bm x,t)-kD\big)$ over a compact state set and one forcing period.
Then $\hat\mu(k)$ is nonincreasing in $k$.
\end{theorem}

\begin{proof}
(i)~Let $V=\delta\bm x^\top P\,\delta\bm x$. Then
$\dot V=\delta\bm x^\top(PA_c+A_c^\top P)\,\delta\bm x\le2\mu_P(A_c)\,V$.
Gr\"onwall's inequality gives
$V(t)\le V(0)\exp\!\big(2\int_0^t\mu_P(A_c(\sigma))\,d\sigma\big)$. The norms
$\|\cdot\|$ and $V^{1/2}$ are equivalent because $P$ is constant. Taking
$t^{-1}\log$ and the upper limit gives the claim. (ii)~For fixed $(\bm x,t)$ and
$k_2>k_1$, the symmetric part at $k_2$ equals that at $k_1$ minus the
positive-semidefinite matrix $(k_2-k_1)D$. Weyl's inequality gives a smaller or
equal largest eigenvalue, and the supremum preserves the order.
\end{proof}

Part~(i) does not use $D\succeq0$. It therefore applies to all four
interventions, including rate scaling. Part~(ii) bounds only $\hat\mu(k)$. It does
not make $\lambda_1$ monotone, and Fig.~\ref{fig:sweeps}(a) shows that
$\lambda_1$ is not monotone. The Euclidean measure can be conservative for
oscillator Jacobians, and a weighted metric can certify contraction where the
Euclidean one cannot. Weighted energy functionals of this kind are already used
in classical feedback analyses \cite{chendong1993}. Let $\rho(M)$ denote the largest real part of the eigenvalues of $M$. Define
its time average along the trajectory by
\begin{equation}
\bar\rho=\limsup_{t\to\infty}\frac1t\int_0^t
\rho\big(A_c(\sigma)\big)\,d\sigma.
\label{eq:rhobar}
\end{equation}

\begin{proposition}[Threshold instability]\label{prop:threshold}
(i)~For every constant $P\succ0$ and every real square matrix $M$,
$\mu_P(M)\ge\rho(M)$. Hence $\bar\mu_P\ge\bar\rho$ for every constant $P$.
(ii)~In the FitzHugh--Nagumo model \eqref{eq:fhn1}--\eqref{eq:fhn2} with
$0<b<1$, evaluate the closed-loop Jacobian on the threshold line $v=0$. Under
recovery damping, $\rho>0$ for every $k\ge0$. Under rate scaling, $\rho>0$ if
and only if $\gamma<1/(\varepsilon b)-1$. Under tracking, $\rho>0$ if and only
if $\kappa<1-\varepsilon b$.
\end{proposition}

\begin{proof}
(i)~Let $M\bm\xi=\ell\bm\xi$ with $\bm\xi$ complex
and nonzero, and let $\bm\zeta=P^{1/2}\bm\xi$. Then $B\bm\zeta=\ell\bm\zeta$,
and $\mathrm{Re}\,\ell=\mathrm{Re}(\bm\zeta^\ast B\bm\zeta)/(\bm\zeta^\ast\bm\zeta)
=\bm\zeta^\ast\tfrac12(B+B^\top)\bm\zeta/(\bm\zeta^\ast\bm\zeta)\le\mu_P(M)$.
The asterisk denotes the conjugate transpose. The inequality is classical \cite{desoer1975}. (ii)~On $v=0$ the Jacobian does not depend on $w$. A real
$2\times2$ matrix has $\rho>0$ if its determinant is negative, or if its
determinant and trace are both positive. Under damping the trace is
$1-\varepsilon b-k$ and the determinant is $\varepsilon(1-b)-k$. If
$k<\varepsilon(1-b)$, the determinant is positive and the trace exceeds
$1-\varepsilon>0$. If $k=\varepsilon(1-b)$, one eigenvalue is zero and the other
equals the positive trace. If $k>\varepsilon(1-b)$, the determinant is negative.
Under rate scaling the trace is $1/(1+\gamma)-\varepsilon b$ and the
determinant is $\varepsilon(1-b)/(1+\gamma)>0$, so $\rho>0$ exactly when the
trace is positive. Under tracking the trace is $1-\kappa-\varepsilon b$ and the
determinant is $\varepsilon(1-b+\kappa b)>0$, with the same
conclusion.
\end{proof}

Proposition~\ref{prop:threshold} has a direct reading. Every firing event
crosses the line $v=0$. Near that line, recovery damping leaves a positive
frozen growth rate at every gain. When the crossing is transverse, continuity keeps $\rho$ positive on a short time interval around it. Proposition~\ref{prop:threshold}(i) then gives a positive contribution to $\bar\mu_P$ from each such firing event, in every constant metric. Contracting stretches of
the orbit must outweigh these contributions before a certificate can hold.
Fast-variable feedback acts on the unstable direction itself. Above
$\kappa=1-\varepsilon b$ it removes the positive frozen growth rate at
threshold. Rate scaling removes it only above $\gamma=1/(\varepsilon b)-1$, far
outside the firing window. In the Aliev--Panfilov model the $(1,1)$ Jacobian
entry at $u=\alpha$ equals $K\alpha(1-\alpha)-q$, which is positive for small
$q$. Recovery damping leaves it unchanged, rate scaling divides it by
$1+\gamma$, and tracking subtracts $\kappa$. The corresponding eigenvalue
statement is checked numerically, not proved. In the Hindmarsh--Rose model the
$(1,1)$ entry is $-3x^2+2\beta x$. Its largest value is $\beta^2/3$, reached at
$x=\beta/3$. Tracking with $\kappa>\beta^2/3$ makes this entry negative for all
$x$, while recovery damping leaves it unchanged. The autonomous model admits a stronger statement. Recovery damping, rate
scaling, and the uncontrolled flow keep the Hindmarsh--Rose closed loop
autonomous. Tracking does not, because $\bar x_1(t)$ depends on time.

\begin{proposition}[Certificates in autonomous loops]\label{prop:auto}
Let $\dot{\bm x}=\bm G(\bm x)$ with $\bm G$ continuously differentiable, and
let $\bm x(t)$ be a bounded trajectory with Jacobian $A_c(t)$ along it. If
$\bar\mu_P<0$ for some constant $P\succ0$, then $\bm x(t)$ converges to an
equilibrium.
\end{proposition}

\begin{proof}
The velocity $\bm\nu(t)=\dot{\bm x}(t)=\bm G(\bm x(t))$
satisfies $\dot{\bm\nu}=A_c(t)\,\bm\nu$. The Gr\"onwall step in the proof of
Theorem~\ref{thm:bound} gives
$\|\bm\nu(t)\|_P\le\|\bm\nu(0)\|_P\exp\!\big(\int_0^t\mu_P(A_c(\sigma))\,d\sigma\big)$,
with $\|\bm\nu\|_P=(\bm\nu^\top P\bm\nu)^{1/2}$. Since $\bar\mu_P<0$, there are
constants $\Lambda>0$ and $C>0$ with $\|\bm\nu(t)\|_P\le Ce^{-\Lambda t}$ for all $t\ge0$.
The velocity is therefore integrable, and $\bm x(t)$ converges to a point
$\bm x^\ast$. Continuity gives
$\bm G(\bm x^\ast)=\lim_{t\to\infty}\dot{\bm x}(t)=\bm 0$.
\end{proof}

Proposition~\ref{prop:auto} is a trajectory-wise form of the known fact that a
contracting autonomous system converges to an equilibrium \cite{lohmiller1998}.
Applied here, it says that no constant-metric certificate can coexist with
sustained firing under recovery damping or rate scaling in the Hindmarsh--Rose
model. A converging trajectory stops crossing threshold, so a certificate in
such a loop certifies a silent cell. Only a time-varying input such as
tracking can escape this limit. The proposition covers ordinary differential
equations. Pyragas control is a delay system and is not treated here.

Constant metrics are evaluated along each computed trajectory. Define the mean
closed-loop Jacobian over the observation interval by
\begin{equation}
\bar J=\frac{1}{t_b-t_a}\int_{t_a}^{t_b}A_c(t)\,dt.
\label{eq:Jbar}
\end{equation}
When $\bar J$ is Hurwitz, one candidate metric solves
\begin{equation}
\bar J^\top P+P\bar J=-\mathsf I.
\label{eq:lyapmetric}
\end{equation}
Additional candidate metrics come from a multistart Nelder--Mead search over
constant positive-definite metrics, up to an irrelevant scale. The search uses
triangular coordinates of $P^{1/2}$. Let $\mathcal P$ denote the finite set of
candidate metrics evaluated by these procedures. The reported value is
\begin{equation}
\bar\mu_\star=\min_{P\in\mathcal P}\bar\mu_P.
\label{eq:mustar}
\end{equation}
By Proposition~\ref{prop:threshold}(i), $\bar\mu_\star\ge\bar\rho$. The
certificate gain $g_c$ is the smallest scanned gain from which every larger
scanned gain gives $\bar\mu_\star<0$. All certificates here are trajectory-wise.
None is a uniform bound over a state set.

\begin{figure}[htbp]
\centering
\begin{subfigure}[b]{0.32\linewidth}
  \centering
  \includegraphics[width=\linewidth]{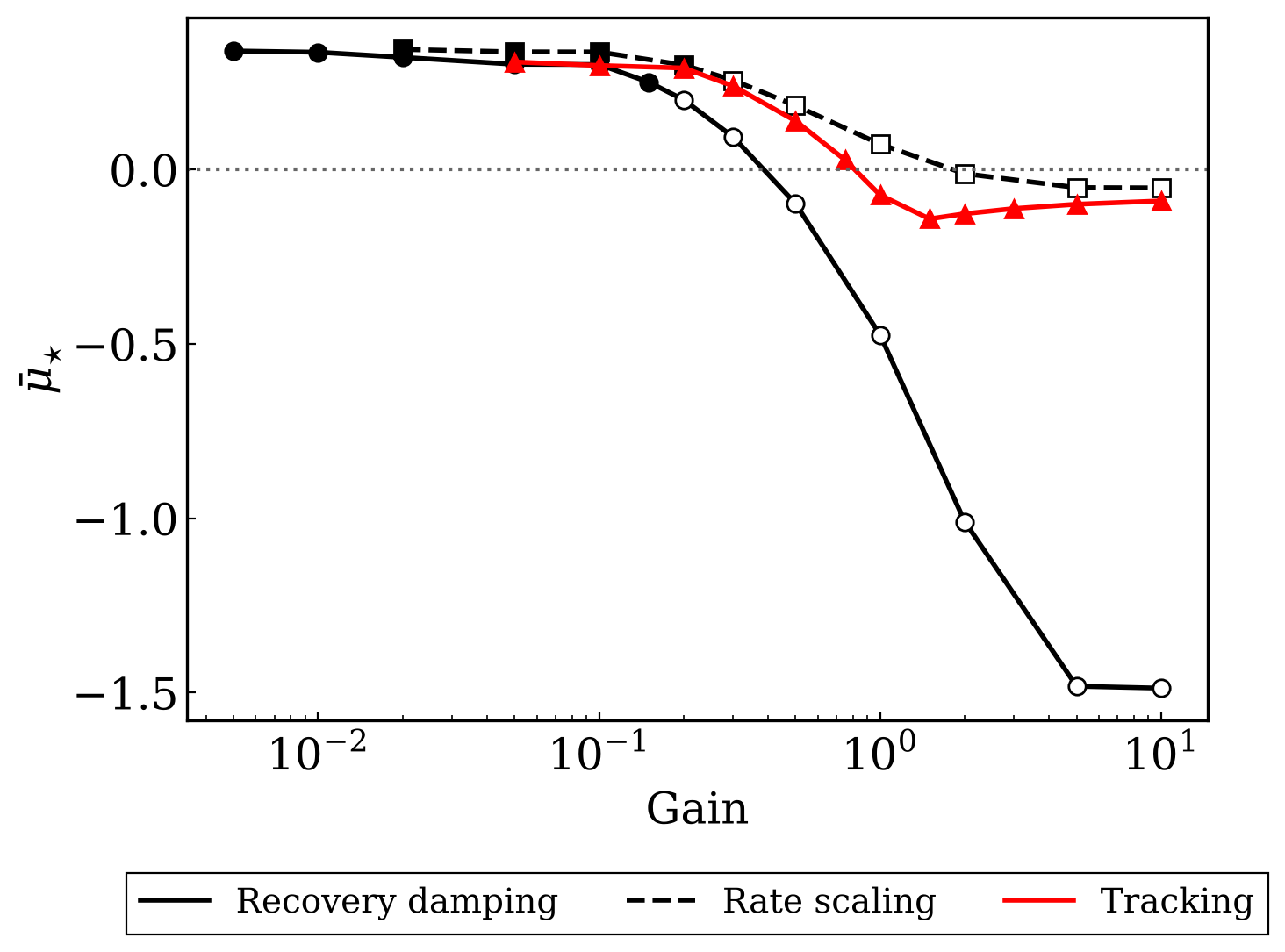}
  \caption{}
\end{subfigure}
\hfill
\begin{subfigure}[b]{0.32\linewidth}
  \centering
  \includegraphics[width=\linewidth]{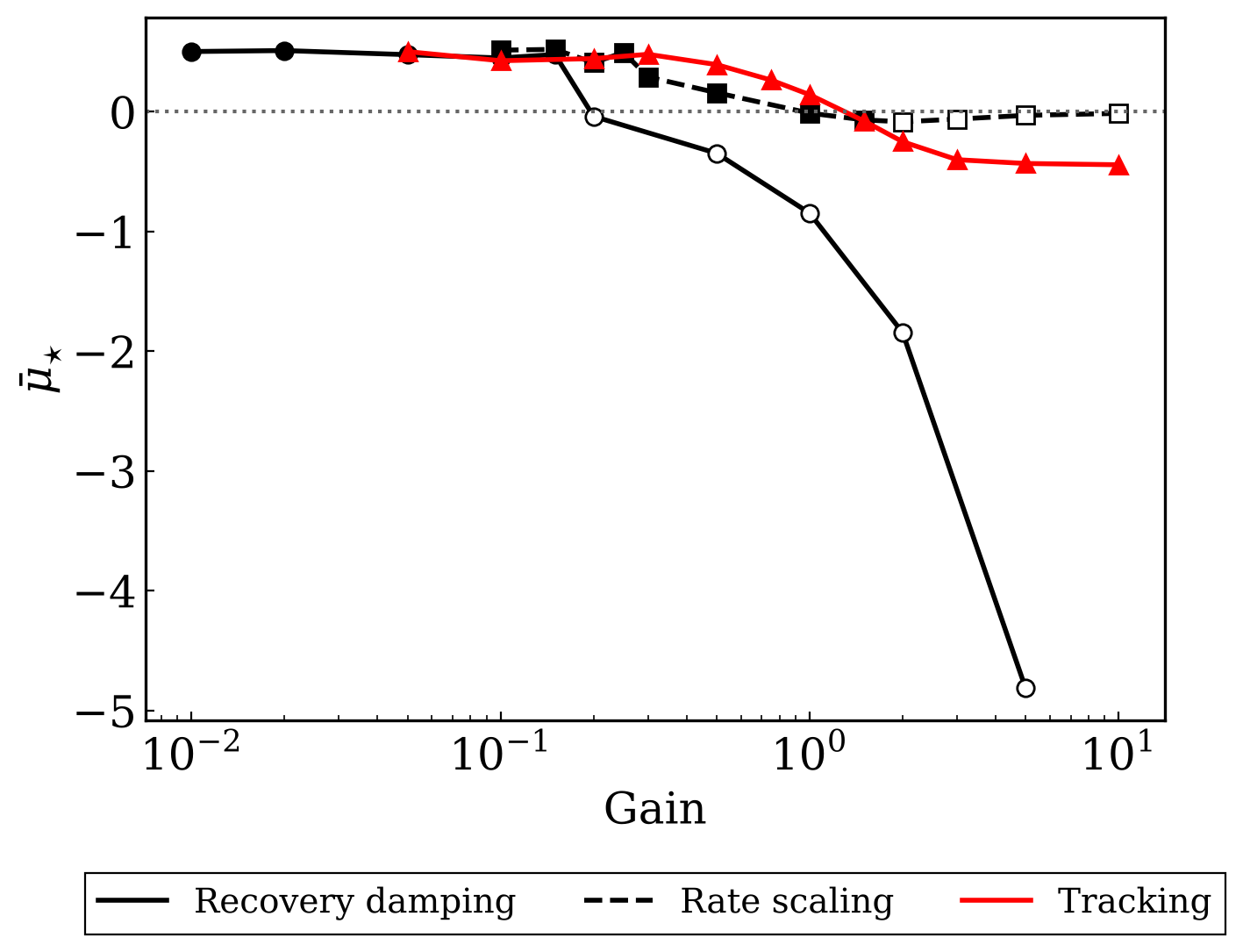}
  \caption{}
\end{subfigure}
\hfill
\begin{subfigure}[b]{0.32\linewidth}
  \centering
  \includegraphics[width=\linewidth]{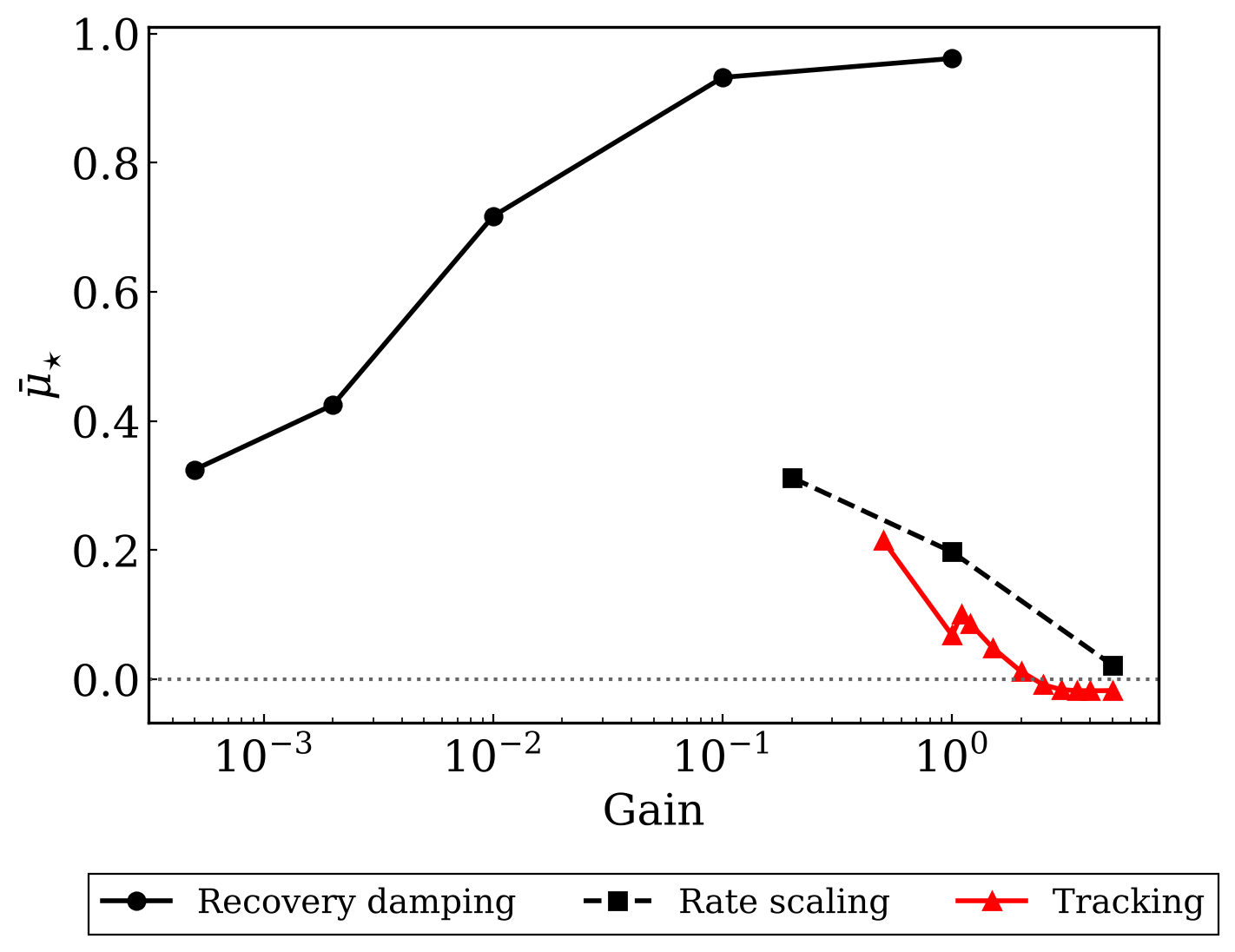}
  \caption{}
\end{subfigure}
\caption{Best constant-metric certificate $\bar\mu_\star$ against gain at the
reference regimes. A negative value certifies $\lambda_1{<}0$ along the computed
trajectory. (a)~FitzHugh--Nagumo model. Filled markers: firing. Open markers:
silent. (b)~Aliev--Panfilov model, with the same markers. (c)~Hindmarsh--Rose
model. The cell fires at every plotted gain.}
\label{fig:cert}
\end{figure}

Fig.~\ref{fig:cert}(a) shows $\bar\mu_\star$ against gain for the
FitzHugh--Nagumo model. Under recovery damping and rate scaling, the curve
crosses zero only where the markers are open. The certificate gain is
$g_c\in(0.39,\,0.40]$ under damping and $g_c\in(1.75,\,1.80]$ under rate
scaling. Both lie above the silencing gains of Table~\ref{tab:thresholds}. The
certificate therefore holds only for a silent cell. Under tracking, the curve
crosses zero at $g_c\in(0.8125,\,0.825]$ while the cell fires every period.
The crossing lies slightly below $\kappa=1-\varepsilon b$.
Removing the threshold instability is therefore not necessary for a
certificate. The orbit crosses the threshold line quickly, so a gain close to
that value suffices.

Fig.~\ref{fig:cert}(b) shows the Aliev--Panfilov model. Under recovery damping,
the certificate holds at the gain where the cell depolarizes, with
$g_c\in(0.1800,\,0.1825]$ equal to the silencing gain. Under tracking, it
holds from $g_c\in(1.300,\,1.325]$ while the cell fires at the uncontrolled
rate with two thirds of the uncontrolled excursion. Under rate scaling, it
holds from $g_c\in(0.925,\,0.950]$ while the cell still crosses $u=\alpha$. At those gains the peak of $u$ stays
below $0.2$, and the excursion is near one fifth of the uncontrolled value.
With a threshold at $u=0.5$, these responses count as silent, and the
certificate again holds only after silencing.

Fig.~\ref{fig:cert}(c) shows the Hindmarsh--Rose model. Under recovery damping
and rate scaling, $\bar\mu_\star$ stays positive at every tested gain, as
Proposition~\ref{prop:auto} requires for a firing cell. Under tracking, $\bar\mu_\star$ becomes negative at $g_c\in(2.0,\,2.5]$ while the
cell fires at the rate of the target orbit. The crossing lies below
$\kappa=\beta^2/3$.

Across the three models, a certificate coexists with substantial firing only
under tracking. Under recovery damping, it holds only after silencing or never.
Under rate scaling, it holds only after silencing or never, except in the
Aliev--Panfilov model. There it holds while the cell still crosses $u=\alpha$,
with an excursion near one fifth of the uncontrolled value.

\section{Discussion}\label{sec:discussion}

The results distinguish regularization that removes excitation from
regularization that stabilizes a firing orbit. Large recovery damping and large
rate scaling remove the chaotic excursion. In the forced models they give
negative exponents and simple certificates, but the cell no longer
fires, or it stays depolarized. Tracking and Pyragas control stabilize a firing
orbit instead. They need knowledge of the target orbit or of its period, and
their steady effort can vanish.

The autonomous model shows that the failure mode depends on the model. Recovery
damping of the Hindmarsh--Rose adaptation variable does not silence the cell at
any scanned gain. It
removes the adaptation that terminates bursts, and the cell switches to dense
tonic firing. A negative or zero exponent is reached, but the firing pattern is
rewritten. The firing-rate ratio exposes this change, while the excursion ratio
does not. Rate scaling lowers the firing rate to occasional threshold crossings on slow waves.
Pyragas control regularizes the cell but settles on an orbit other than the
target, with finite steady effort. The excitation measures are therefore
complementary. No single one captures every failure.

The high-gain recommendation of \cite{chendong1993} aimed to let the feedback
dominate the forcing. In an excitable cell, high gain on the recovery variable
either silences the cell or rewrites its firing pattern. The gain intervals of
Table~\ref{tab:thresholds} give a lower limit on the useful gain. Silencing or a
large change of $S$ gives an upper limit.

The numerical sign of $\lambda_1$ was audited for the Aliev--Panfilov model at
$F=0.40$ and $\omega=1.8$. The baseline run uses the step $T/350$, a
renormalization interval of 50 steps, a transient of 57 forcing periods, an
observation window of 430 periods, and the initial condition $(u,q)=(0.1,0.1)$.
Each audit run changes one of these choices. The initial condition is also set
to $(0.5,0.2)$ and $(0.05,1.0)$. The step ranges from $T/175$ to $T/700$, the
renormalization interval from 25 to 100 steps, the transient from 29 to 115
periods, and the observation window from 286 to 716 periods. The uncontrolled
exponent stays between $+0.1005$ and $+0.1325$, around the baseline value
$+0.1111$. The exponent at $\gamma=5$ equals $-0.0414$ in every run. The
sign is therefore unchanged under all these variations. The magnitude of the positive exponent varies by up to about twenty percent across
these variations. That spread is the finite-time variability of a chaotic
exponent. Nine combinations of forcing errors of $-15$, $0$, and $+15$ percent
in $F$ and $\omega$ keep the exponent at $\gamma=5$ between $-0.0704$ and
$-0.0284$. The same errors in $A$ and $\omega$,
applied at $A=1.30$, keep the FitzHugh--Nagumo exponent negative under damping
with $k=5$, over the range from $-1.730$ to $-0.448$. Both tests concern
high-gain, silenced regimes. They are sign checks, not proof that the firing
window is robust to parameter uncertainty.

The study has limits. All results are numerical and cover three low-order
models at one operating point each. Each threshold is known only to within one
step of its fine scan, and the exponent can change sign more than once inside a
step. The excitation index depends on the chosen threshold, as the
Aliev--Panfilov rate-scaling entry of Table~\ref{tab:thresholds} shows. The
Hindmarsh--Rose regularity test relies on a tolerance for the zero exponent of
a periodic orbit. The certificates are trajectory-wise and use constant metrics
only. A time-varying metric could certify more. Proposition~\ref{prop:threshold}(ii)
is proved for the FitzHugh--Nagumo model only. Tracking needs the target orbit,
which an application would have to estimate from measurements. Pyragas control
needs the period. None of the interventions is proposed as a therapy.

\section{Conclusion}\label{sec:conclusion}

Chaos control in an excitable oscillator has two goals, regularity and
retained firing. A negative Lyapunov exponent measures only the first.
Excitation measures placed next to the exponent separate the two outcomes on
the same chaotic regimes of three models. In the FitzHugh--Nagumo model,
recovery damping and rate scaling leave a window of gains that suppresses chaos
and keeps firing. In the Aliev--Panfilov model, only rate scaling leaves such a
window. Recovery damping removes chaos and firing at the same scanned gain.
Larger gains fall outside the window. In the autonomous model, recovery damping
keeps the cell firing but multiplies its firing rate. Orbit tracking suppresses
chaos and keeps firing in all three models. Pyragas control also keeps firing,
but in the Hindmarsh--Rose model it settles on an orbit other than the target.
A classical matrix-measure bound, applied at the firing threshold and to
autonomous loops, explains this split. A constant-metric certificate coexists
with substantial firing only under tracking, the one intervention that
subtracts a gain from the fast-direction entry of the Jacobian. Gain selection
for an excitable cell therefore needs an excitation measure as well as the
exponent.

\section*{Acknowledgments}

AI tools were used for editorial assistance, proofreading, and language refinement. All theoretical developments, derivations and figures are the authors' work.

\section*{Data Availability}

No external data sets were analyzed. 

\bibliographystyle{ws-ijbc}
\bibliography{refs}

\end{document}